# ULTRAVIOLET DIVERGENCES IN EXTENDED SUPERGRAVITY *)

M.J. Duff **)

CERN -- Geneva

## ABSTRACT

Recent calculations have confirmed the belief that the ultraviolet behaviour of N-extended supergravities improves with increasing N. We give a comprehensive review.

## CONTENTS

---

*) Lectures delivered at the Spring School on Supergravity, ICTP, Trieste, April 1981, to Appear in Supergravity 81, edited by S. Ferrara and J.G. Taylor and published by Cambridge University Press.

**) Permanent address : Blackett Laboratory, Imperial College, London.

# 1 INTRODUCTION

## 1.1 Quantum gravity and grand unification

These lectures are about the ultraviolet problem in gravity, but I would like to begin with some thoughts on the grand unification of strong, weak, and electromagnetic interactions. Here, the question is "How many of the observed, and yet to be observed, particles are truly elementary and what are their quantum numbers?" The answer might lie in SU(5) or SO(10) or maybe some more primitive theory of "preons", but it remains a mystery why Nature should single out one symmetry from among the many mathematical possibilities. To begin with, how does one count the number of particle species? Well, there is a sense in which gravity does just that. Consider, for example, closed-loop corrections to the graviton self-energy. By the Equivalence Principle, gravity couples to everything (including itself) with equal strength. Thus no matter whether the particle going round the loop is a quark, a W boson, a photon, or whatever; each contributes to the self-energy with the same order of magnitude. For consistency, therefore, one must include <u>all</u> the elementary particles irrespective of their masses or internal quantum numbers. In this way gravity cares, in a way which no other force does, just how many particles there are. Thus it is not inconceivable that gravity may have something to say about the spectrum of the elementary particles.

Now let us turn to the other side of the coin, to the problem of constructing a consistent quantum theory of gravitation. Here, the question is "How does one make sense of a theory which, by power-counting at least, is non-renormalizable?" As we shall recall in Section 2, the superficial degree of divergence of a Feynman graph is given by $D = (d - 2)L + 2$, where d is the dimension of space-time and L the number of closed loops. For $d = 4$, this increases with increasing loop order and leads to a non-renormalizable theory. One way out might be to couple gravity to matter fields and to look for a mutual cancellation of ultraviolet divergences. Note that this represents a shift in philosophy away from conventional renormalizability, and away also from the old ideas of an ultraviolet cut-off at the Planck length. Rather one hopes that on-shell S-matrix elements will be finite order by order in perturabtion theory. It should be clear, however, that if this idea is to work at all it could work only for some very special assignment of masses, spins, and internal quantum number of the matter fields. Thus it is not inconceivable that the spectrum of the elementary particles may, in its turn, have something to say about gravity.

With these premisses, the most economical conclusion is that the problem of constructing a grand unified strong electro-weak theory and the problem of constructing a quantum theory of gravity are really one and the same problem: only with the right elementary particles will the theory be finite; finiteness determines the right elementary particles. And before dismissing the whole idea as being too fantastic, one should recall, as we shall in Section 2, some of the alternative proposals for solving the renormalizability problem in gravity in comparison to which, one could argue, the present idea is rather conservative. [After all, we are already used to the anomaly-free criterion in grand unified theories (GUTs), whereby the absence of certain divergences is invoked to restrict the allowed numbers of quarks and leptons.]

This idea is not new and was pursued before the discovery of supergravity, albeit without much success. First it was realized from general positivity arguments that contributions to the graviton self-energy from particles of spin 0, $\frac{1}{2}$, and 1 entered not only with the same order of magnitude but also with the same sign (Capper et al. 1974; Capper & Duff 1974 a; Deser & van Nieuwenhuizen 1974 b.) So hopes of infinity cancellations in off-shell Green's functions seemed hopeless. More promising was the idea of finite on-shell S-matrix elements, but although pure gravity was found to be all right at one-loop ('t Hooft & Veltman 1974), gravity coupled to various combinations of spins 0, $\frac{1}{2}$ and 1 gave infinite results. (A list of references is given in Section 2.) And the prospect for higher loop order, with or without matter, seemed even bleaker. Even at the time, one was aware of two shortcomings in this approach. First, in no sense was there a unification of gravity and matter; one simply picked one's favourite matter theory and only afterwards grafted on the gravity. Secondly, it was completely hit-and-miss in its attempts to find the magic combinations of matter fields; there was simply no guiding principle. This state of affairs was reviewed by Duff (1975) and Deser (1975) in the Proceedings of the 1974 Oxford Quantum Gravity Conference. With characteristic foresight, Salam (1975) pointed out in the same volume a third possible shortcoming: the neglect of spin $\frac{3}{2}$.

### 1.2 Supergravity

By consistently coupling spin-2 gravitons to spin-$\frac{3}{2}$ "gravitinos" simple (N = 1) supergravity (Freedman et al. 1976; Deser & Zumino 1976) provided the first example of a gravity-matter system yielding finite on-shell S-matrix elements at one loop (Grisaru et al. 1976). But it did much more. Here, for the first time, was the dreamed-of unification: gravity and matter as merely different components of the same symmetry multiplet, and Bose-Fermi symmetry as an obvious candidate for that missing guiding principle. These features become even more striking when one considers the extended ($1 < N \leq 8$) supergravities and especially the $N = 8$ theory which combines one spin-2, eight spin-$^3/_2$, twenty-eight spin-1, fifty-six spin-$^1/_2$, and seventy spin-0 particles in one supermultiplet. Although the internal symmetry assignments prevent these particles from being identified with those observed at present energies, they might possibly be the preons from which bound-state quarks and so on are formed (Ellis et al. 1980; Derendinger et al. 1981). Now conventional model-building via preons is prejudiced by the economical requirement that the number of pre-particles be small; a criterion obviously violated by $N = 8$ supergravity. However, if one thinks in terms of pre-fields rather than pre-particles then it is the ultimate in economy since there is but one single superfield.

Naturally, two questions now arise (a) Is there a finite theory of extended supergravity? (b) Does it describe the right particles? These lectures will summarize what we know in response to (a). The reader is also referred to the lectures by Dr. Kallosh at this school. Other recent reviews on ultraviolet divergences may be found in Weinberg (1979) and van Niewenhuizen (1981 a). As described in detail by Dr. Kallosh, and summarized here in Section 2, it has now been established that supergravity theories are on-shell finite both at one and two loops, but that superinvariants exist as possible counterterms at three loops and beyond. [We are assuming here (a) trivial space-time topology, (b) no cosmological constant, and (c) no breaking of supersymmetry. We shall deal with cases (a) and (b) later on. Case (c) is discussed by van Nieuwenhuizen (1981 b).] As Kallosh explains, moreover, going to higher N (e.g. $N = 8$) does not avoid the problem of invariants as potentially dangerous counterterms. The only hope remaining, it seems, is that the coefficient of such invariants must vanish in the counterterm. (And here we remind the reader that, despite great efforts in quantum gravity, no explicit calculations yet exist beyond one loop.) This might seem like clutching at straws were it not for the fact that we already have concrete examples in supersymmetry and supergravity where this non-renormalization phenomenon actually occurs!

The first example, discussed in Section 3, concerns extended supergravity with local SO(N) invariance. When the internal SO(N) symmetry is gauged, the usual arguments for one-loop finiteness cease to apply because of the appearance of a cosmological constant related to the gauge coupling constant e. Indeed, for $N \leq 4$ one finds that infinite renormalizations are required. Remarkably, the particle content of theories with $N > 4$ results in a cancellation of these infinities implying, in particular, a vanishing one-loop $\beta(e)$ function (Christensen et al. 1980). This is reminiscent of the vanishing $\beta$-function in supersymmetric Yang-Mills theories for $N > 2$, which is now known to hold to at least three-loop order (Avdeev et al. 1980; Grisaru et al. 1980; Caswell & Zanon 1980). The point I wish to emphasize is that in both cases candidate counterterms do exist but nevertheless appear with zero coefficient.

The purpose of Section 3 will be to show how, to one loop at least, these otherwise "miraculous" cancellations have a common explanation in terms of certain "spin-moment sum rules" (Curtwright 1981). These sum rules provide at last concrete evidence that higher N means better ultraviolet behaviour, as had long been hoped for. In showing how these spin sum rules are relevant to ultraviolet divergences and anomalies, we shall revive some earlier work of Christensen & Duff (1978 a, 1979) on counterterms, axial anomalies, and conformal anomalies for fields of arbitrary spin.

Another point of technical interest concerns the cosmological constant $\Lambda$ related to the Yang-Mills coupling constant e of the gauged extended supergravities by $\kappa^2\Lambda = -6e^2$, where $\kappa^2 = 16\pi G$ and G is Newton's constant. To calculate the $\beta(e)$ function, therefore, one may either (a) compute the usual charge renormalization effects, i.e. the coefficient of the Yang-Mills Tr $\sqrt{g}\, F^{\mu\nu}F_{\mu\nu}$ counterterm, which receives contributions from spins 0, $\frac{1}{2}$, 1, and $\frac{3}{2}$ but not 2, since the graviton is a singlet, or (b) compute the cosmological renormalization i.e. the coefficient of the $\sqrt{g}$ counterterm, which receives contributions from all spins including gravity. By supersymmetry, one coefficient determines the other. (Incidentally, it is amusing to note that, whether or not one believes in supergravity, this enables one to deduce the magnitude of graviton loop effects in pure gravity from knowledge of flat-space Yang-Mills theories.) In Section 3 we shall calculate $\beta(e)$ both ways and demonstrate their equivalence. In order to carry out method (b), however, one must first understand how to handle quantum effects of gravity with a non-vanishing cosmological constant (Christensen & Duff 1980), which we shall also briefly describe.

As we have discussed, it is our purpose in these lectures to concentrate on those ultraviolet properties which are peculiar to particular values of N, rather than dwell on those common to all N. In Section 4 we return to ordinary (i.e. ungauged) supergravity and examine another aspect of N dependence, namely anomalies and topological counterterms. In gravity theories, the anomalous contribution to the trace of the effective energy momentum tensor receives a one-loop contribution proportional (on-shell) to $\varepsilon^{\mu\nu\rho\sigma}R_{\rho\sigma\alpha\beta}\varepsilon^{\alpha\beta\gamma\delta}R_{\mu\nu\gamma\delta}$ which, when integrated over all space, yields a topological invariant: the Euler number $\chi$ (Duff 1977). Now anomalies arise because of divergences, and there is a corresponding counterterm proportional to $\chi$ which is non-zero in spaces with non-trivial topology. The numerical coefficient of this anomaly, call it A, has been calculated for extended supergravity theories and found to be non-vanishing, and non-integer, for N = 1 and 2 but equal to an integer, A = 3-N, for $N \geq 3$ (Christensen & Duff 1978 a; Christensen et al. 1980). These calculations assume the usual field representation assignments for each spin. Recently, however, it was shown that A depends not only on spin but also on choice of representation (Duff & van Nieuwenhuizen 1980). Thus, contrary to naïve expectations, the gauge theory of a rank-two antisymmetric tensor field $\phi_{\mu\nu}$ is not equivalent to a scalar $\phi$, even though both describe spin-0. Similarly, the gauge theory of a rank-three antisymmetric field $\phi_{\mu\nu\rho}$ is not equivalent to nothing.

These results might be of only academic interest were it not for the appearance of such unusual representations in the auxiliary fields of simple supergravity and in the versions of extended supergravity obtained via dimensional reduction. For example, N = 1 supergravity in d = 11 dimensions contains a rank-three field $\phi_{\mu\nu\rho}$. After dimensional reduction to d = 4 dimensions, one obtains an N = 8 theory, not with seventy scalars, but with sixty-three $\phi$, seven $\phi_{\mu\nu}$, and one $\phi_{\mu\nu\rho}$. Remarkably, the A coefficient for N = 8 with these representations now vanishes! A similar phenomenon happens for the d = 4, N = 4 theory obtained from d = 10, N = 1, where the two spin-0 fields appear as one $\phi$ and one $\phi_{\mu\nu}$. So now we have A = 0 not only for N = 3, but also N = 4 and N = 7, 8. By extrapolating the representation assignments to N = 5 and 6, one can arrange for A = 0 for all $N \geq 3$ (Duff 1981 b,c; Nicolai & Townsend 1981). The derivation of these results has been discussed at length elsewhere (Duff 1981 c) and so in Section 4, I shall instead concentrate on their interpretation within the framework of superfield quantization and superfield Feynman graphs (Grisaru & Siegel 1981 a). In particular, we shall note that of the two kinds of superfield, chiral and non-chiral, only closed loops of chiral superfields contribute to the A coefficient. By analysing the extended theories in terms of N = 1 superfields, therefore, one can explain the absence of anomalies and the finiteness by noting that the net number of chiral superfields (i.e. physical minus ghost) is zero for $N \geq 3$.

We also discuss in Section 4 how anomaly coefficients can change not only with a change of physical and/or auxiliary field assignments but also with a change of boundary conditions.

Since these lectures were delivered, there have been several interesting new developments in the subject. These are summarized in Section 5.

### 1.3 Kaluza-Klein?

Finally, I should mention that there has recently been a renewed interest in higher dimensional theories of the Kaluza-Klein type. Here one begins with a gravity theory on $M \times B$, where M is four-dimensional space-time and B is some compact space, and ends up with a gravity-Yang-Mills theory on M with a gauge group determined by the symmetries of B. There are two reasons why N = 1 supergravity in d = 11 is particularly interesting in this connection. First, as pointed out by Witten (1981), eleven dimensions is both the minimum number to accommodate $SU(3) \times SU(2) \times U(1)$ as a gauge group and the maximum number allowed by supersymmetry. (Witten's paper contains many other interesting results.) Secondly, as pointed out by Freund and Rubin (1980), preferential compactification to four (or seven) dimensions is found to occur dynamically. This is because a rank-three gauge potential $\phi_{\mu\nu\rho}$ can give rise to a cosmological constant (Duff & van Nieuwenhuizen 1980; Aurelia et al. 1980) and its appearance in d = 11 supergravity is just what is required to make $M \times B$ a solution of the field equations with B compact with d = 7 and M non-compact with d = 4. Note that in this Kaluza-Klein picture, the extra dimensions must be taken seriously. This is to be contrasted with the dimensional reduction discussed earlier, which is merely a cunning device for determining the N = 8 Lagrangian in d = 4, and corresponds to discarding all but the massless modes.

In this Kaluza-Klein picture, therefore, the question is not whether N = 8 supergravity is finite in four dimensions, but whether N = 1 supergravity is finite in eleven dimensions!

In d = 11 the degree of divergence is 9L + 2, which is of course worse than in d = 4. Remember, however, that we are not looking for power-counting renormalizability but finiteness due to a cancellation of ultraviolet divergences. *A priori*, this seems to me just as likely in d = 11 as in d = 4. Moreover, in odd dimensions, gravity theories are automatically finite at odd loop order since there are no invariants formed from the metric involving an odd number of derivatives. In this respect, we are already half-way there!

In any event, the lectures presented here are based on the prejudice that we live in four dimensions. Ultraviolet divergences in Kaluza-Klein theories will be treated elsewhere (Duff & Toms 1981).

## 2 REVIEW OF RENORMALIZABILITY PROBLEM

### 2.1 Pure gravity

Consider the Lagrangian for pure gravity with zero cosmological constant

$$\mathcal{L} = -\frac{1}{16\pi G}\sqrt{g}\,R\ . \tag{2.1}$$

Since the curvature scalar R involves two derivatives of the metric, the corresponding momentum-space vertex functions will behave like $p^2$, and the propagator like $1/p^2$. In d dimensions each loop integral will contribute $p^d$, so with L loops, V vertices, and P internal lines, the superficial degree of divergence of a Feynman diagram is given by

$$D = dL + 2V - 2P\ . \tag{2.2}$$

Combined with the topological relation

$$L = 1 - V + P \tag{2.3}$$

this yields

$$D = (d-2)\,L + 2\ . \tag{2.4}$$

Note that D does not depend on the number of external lines. The crucial point is that D increases with increasing loop order for $d > 2$ and leads to a non-renormalizable theory. For $d = 2$, Eq. (2.1) ceases to have any dynamical content since $\mathcal{L}$ is a total derivative. Let us see what this means in practice for $d = 4$ within the framework of some specific regularization scheme. We shall use dimensional regularization which means working in $4 + \varepsilon$ dimensions and then letting $\varepsilon \to 0$ at the end. At one loop, $D = 4$, which means we expect the one-loop counterterms $\mathcal{L}_{(1)}$ to depend on four derivatives or less. On dimensional grounds, the only generally covariant scalars available are $R_{\mu\nu\rho\sigma}R^{\mu\nu\rho\sigma}$, $R_{\mu\nu}R^{\mu\nu}$, and $R^2$. Hence

$$\mathcal{L}_{(1)} = \frac{1}{\varepsilon}\sqrt{g}\left[\alpha R_{\mu\nu\rho\sigma}R^{\mu\nu\rho\sigma} + \beta R_{\mu\nu}R^{\mu\nu} + \gamma R^2\right]. \tag{2.5}$$

We are assuming here that the background field method has been employed, (see Section 3) so that the counterterm depends only on the background metric and not on the gravitons or ghosts going round the loop. The constant $\alpha$ is independent of one's choice of gauge for the quantum graviton field but the constants $\beta$ and $\gamma$ are not. This corresponds to the fact that off-mass-shell Green's functions may be gauge-dependent. It is the on-shell S-matrix elements which correspond to the gauge-invariant physics. Within the framework of the background field method, putting the external lines on mass shell corresponds to using the classical equations of motion for the background field, i.e. $R_{\mu\nu} = 0$. Before doing so, however, we first note that the combination $R_{\mu\nu\rho\sigma}R^{\mu\nu\rho\sigma} - 4R_{\mu\nu}R^{\mu\nu} + R^2$ is a total divergence and its integral over all space may be neglected provided space-time has trivial topology, which we assume for the moment (otherwise it yields a topological invariant, the Euler number, which takes on integer values in spaces with non-trivial topology: see Section 4). Consequently $\int d^4x\, \mathcal{L}_{(1)}$ vanishes on-shell, and hence at one loop order on-shell S-matrix elements are actually finite.

The real problem arises when we go beyond one loop. At two loops, for example, D = 6 and in addition to terms vanishing with the field equations we anticipate terms like $\sqrt{g}\, R_{\mu\nu}{}^{\alpha\beta}R_{\alpha\beta}{}^{\gamma\delta}R_{\gamma\delta}{}^{\mu\nu}$ which do not. Consequently divergences would survive even for on-shell S-matrix elements which can be removed only by counterterms of a kind not present in the original Lagrangian. In general, we anticipate an infinite number of distinct counterterms and correspondingly an infinite number of undetermined parameters: the disaster of non-renormalizability.

In the literature one may find several possible responses to this disaster:

a) Quantum gravity makes no sense. In other words, one should quantize all matter fields but keep the gravitational field classical. See, for example, Kibble (1981).

b) Quantum gravity is all right, but the problem is perturbation theory, i.e. sum all graphs, or some appropriate subset of all graphs, in the hope that the result will be finite and unambiguous. This was the idea behind the old non-polynomial Lagrangian approach (Isham et al. 1971). Again, the problem was not so much in obtaining a finite answer, but in obtaining a finite answer which was unambiguous.

c) Perhaps renormalizability is, after all, the wrong criterion. A recent proposal which falls into this category is that of Weinberg's "Asymptotic Safety" programme (Weinberg 1979).

d) Modify Einstein's Lagrangian to include terms quadratic in the curvature and hence depending on four derivatives of the metric (DeWitt 1965; Stelle 1977). Now the dominant behaviour of the vertices is $p^4$ and that of the propagators $1/p^4$. Hence, in four dimensions $D = 4L + 4V - 4P = 4$. Thus no counterterms are required beyond those of the kind already present in the original Lagrangian. However, renormalizability has been bought at the expense of apparent lack of unitarity, since four derivative theories contain unphysical ghost-poles in the propagators. Various arguments have been put forward to circumvent this unitarity problem but none with complete success. Summing bubble graphs, for example, seems to lead to a lack of causality (Tomboulis 1980); whereas propagating torsion theories can be either unitary or renormalizable but not both simultaneously (Sezgin & van Nieuwenhuizen 1980 a). It remains unclear whether this apparent lack of unitarity is an artefact of one's approximation scheme or whether it would disappear in the exact theory. The reader is referred to the literature (Julve & Tonin 1978; Salam & Strathdee 1978 a; Weinberg 1979, Fradkin & Tseytlin 1981 a,b,c; Christensen 1981). Under the category of fourth-order theories we should also include the "induced gravity" approach, initiated by Sakharov (1968), whereby the Einstein Lagrangian, plus fourth-order terms, is induced by quantum matter effects. A review of induced gravity theories has recently been given by Adler (1981).

e) The problem is not with Einstein's theory *per se*, nor with perturbation theory, but with the failure to include precisely the correct set of matter fields. In other words there exists a, possible unique, choice of matter fields for which a mutual cancellation of infinities occurs leading to finite on-shell S-matrix elements order by order in perturbation theory.

It is not my intention here to attack or defend proposals (a) to (d) except to repeat the comment made in the Introduction that, according to current ideas in quantum field theory, proposal (e), though bold, seems to me conservative by comparison. My own objections to semi-classical approaches may be found elsewhere (Duff 1981 a).

## 2.2 Gravity plus matter

Whether or not we adopt viewpoint (e) above, it is natural to ask what happens when coupling to matter is allowed. Again one can write down the most general one-loop counterterm consistent with dimensional analysis, general covariance, and whatever other symmetries are present in the theory. Thus, in addition to $\sqrt{g}\ R_{\mu\nu}R^{\mu\nu}$ and $\sqrt{g}\ R^2$ terms, one might expect $\kappa^2\sqrt{g}\ R_{\mu\nu}T^{\mu\nu}$, $\kappa^2\sqrt{g}\ RT^{\mu}{}_{\mu}$ terms or $\kappa^4\sqrt{g}\ T_{\mu\nu}T^{\mu\nu}$, $\kappa^4\sqrt{g}\ T^{\mu}{}_{\mu}T^{\nu}{}_{\nu}$ terms, where $T_{\mu\nu}$ is the energy-momentum tensor of the matter fields. Other, more complicated terms involving matter fields will also be present in general. Unlike pure gravity, however, explicit calculations are now required to fix the numerical coefficients. These calculations have now been carried out for various gravity-matter systems ('t Hooft & Veltman 1974; Deser & van Nieuwenhuizen, 1974 a,b; Deser et al. 1974; Nouri-Mogadom & Taylor 1975; Sezgin & van Nieuwenhuizen 1980 b; Duff & van Nieuwenhuizen 1980; Van Proyen 1980; Barvinsky & Vilkovisky 1981). In general, all terms which are allowed by the symmetry appear with non-vanishing coefficients. If the matter fields are massive (m = 0), moreover, then one also acquires new divergences like $m^4\sqrt{g}$ and $m^2\sqrt{g}\ R$. (Such terms disappear in the massless limit if, as we are assuming, one employs a regularization scheme with a dimensionless regularizing parameter. With a dimensionful cut-off $\Lambda$, terms like $\Lambda^4\sqrt{g}$ and $\Lambda^2\sqrt{g}\ R$ survive even for massless theories.) Occasionally, however, there are some surprises when *a priori* allowed counterterms do not appear, for example the vanishing of $\sqrt{g}\ R_{\mu\nu\rho\sigma}F^{\mu\nu}F^{\rho\sigma}$ in Einstein-Maxwell theory (Deser & van Nieuwenhuizen 1974 b) or the finiteness of the gravitational modification to the anomalous magnetic moment of the electron in gravity-modified QED (Berends & Gastmans 1975). These can be explained by invoking some non-obvious symmetry (like duality invariance) or else by embedding in supergravity (Deser 1981; van Proyen 1980).

The crucial question, of course, is whether such one-loop counterterms vanish on-shell. By "on-shell" we now mean on using the non-vacuum Einstein equations $R_{\mu\nu} - \frac{1}{2}\ g_{\mu\nu}R = \kappa^2 T_{\mu\nu}$ plus the equations of motion for the matter fields. For all combinations and representations of fields with spin 0, $\frac{1}{2}$, and 1 which have been tried to date, the answer is no! Thus otherwise "respectable" theories like QED, QCD, Weinberg-Salam, or GUTS (all of which were obtained by requiring renormalizability) cease to make sense when gravity is present.

It is perhaps hardly surprising that the coupling of a renormalizable theory like QED to a non-renormalizable theory like gravity, leads to divergent results. Should one, in order to obtain a finite theory, couple gravity to another non-renormalizable theory? An example would be gravity plus non-linear $\sigma$-model (Duff & Goldthorpe 1981). This theory is not finite, either. [The motivation for the calculation of one-loop counterterms was rather to demonstrate the inconsistency of quantum field theory in curved space-time (Duff 1981 a).] However, it is interesting to note that the one-loop counterterms for the $\sigma$-model with coupling constant $\kappa'$ are already in flat space of the form $\kappa'^4 T_{\mu\nu}T^{\mu\nu}$ and $\kappa'^4 T^\mu{}_\mu T^\nu{}_\nu$, thus increasing the probability of cancellations with the previously mentioned curved-space counterterms if $\kappa'$ is chosen to equal $\kappa$. So this may be a step in the right direction, and such couplings do in fact occur in extended supergravity. It is to supergravity that we now turn.

## 2.3 Simple supergravity

The Lagrangian for simple supergravity (Freedman et al. 1976; Deser & Zumino 1976)

$$\mathcal{L} = -\frac{1}{2\kappa^2}\, eR + \frac{1}{2}\,\varepsilon^{\mu\nu\rho\sigma}\bar{\psi}_\mu\gamma_5\gamma_\nu D_\rho\psi_\sigma + \frac{e}{3}\left(S^2 + P^2 - b^\mu b_\mu\right) \qquad (2.6)$$

describes the coupling of gravity $e_\mu{}^a$ to a single spin-$\frac{3}{2}$ Majorana spinor $\psi_\mu$ and does not fall into the category of gravity-matter systems discussed previously. We have also included in Eq. (2.6) the minimal set of auxiliary fields S, P, and $b_\mu$ which vanish on-shell but are necessary for the closure of the supersymmetry algebra off-shell (Stelle & West 1978; Ferrara & van Nieuwenhuizen 1978 a). We shall return to these in a moment. The first quantum-loop calculations in supergravity were in fact carried out before these auxiliary fields were known. As is by now well-known, supergravity provided the first example of a gravity-matter system which was one-loop finite on-shell (Grisaru et al. 1976). In fact, one can show that the one-loop counterterm vanishes when both the Einstein and Rarita-Schwinger field equations are satisfied because it takes the symbolic form

$$\mathcal{L}_{(1)} = \frac{1}{\varepsilon}\,(\text{field equations})^2 \; . \qquad (2.7)$$

(Once again, we have ignored topological effects. See Section 4.) The reason why supergravity succeeded where all other gravity-matter couplings had failed was not an accident, of course, but due to the extra symmetry of Eq. (2.6), i.e. the supersymmetry: there are no supersymmetric quantities of the right dimension which can contribute to $\mathcal{L}_{(1)}$ without also vanishing on-shell. In this sense, it matches pure gravity.

Remarkably, however, supergravity goes one better than pure gravity in being finite on-shell at two loops! (Grisaru 1977). There are no supersymmetric quantities of the right dimension which can contribute to $\mathcal{L}_{(2)}$ without also vanishing on-shell. In other words: a) there is no fermionic partner of the $(\text{Riemann})^3$ invariant discussed previously; b) no new fermionic invariants, without bosonic partners, appear.

Unfortunately, this pattern breaks down at three loop order. As was shown by Deser et al. (1977) an on-shell superinvariant exists, at least at the linearized level, which might act as a three-loop counterterm. It corresponds, in fact, to the supersymmetric completion of the square of the Bel-Robinson tensor. Thus invariance arguments alone are not sufficient to rule out counterterms in simple supergravity and the case for finiteness, although not lost, remains unproved.

It remains, of course, to confirm that the three-loop invariant survives at the non-linear level and to investigate four loops and beyond. This requires a more systematic approach, using either the tensor calculus (Ferrara & van Nieuwenhuizen 1978 b) or else superspace. For an introduction to superspace, see Salam & Strathdee (1974) and Strathdee (this volume). To analyse higher loop counterterms, the Wess-Zumino superfields (Wess & Zumino 1977) prove most useful, because all supertorsions and supercurvatures can be expressed in terms of just three superfields

$$R\ , \quad G_{\alpha\dot{\alpha}}\ , \quad W_{\alpha\beta\gamma}\ ,$$

where we have used two-component spinor notation. All the component field equations (Einstein, Rarita-Schwinger, and auxiliary field) are contained in

$$R = 0\ , \quad G_{\alpha\dot{\alpha}} = 0\ . \tag{2.8}$$

$W_{\alpha\beta\gamma}$, on the other hand, survives on shell and is given by

$$W_{\alpha\beta\gamma} \sim F_{\alpha\beta\gamma} + C_{\alpha\beta\gamma\delta}\theta^{\delta}\ , \tag{2.9}$$

where $F_{\alpha\beta\gamma}$ is the spin-$\frac{3}{2}$ field strength, $C_{\alpha\beta\gamma\delta}$ is the Weyl tensor, and $\theta^{\delta}$ are the fermionic superspace coordinates. Invariants can be built only from these fields and their covariant derivatives. The structure of the one-loop counterterm Eq. (2.7) and both one- and two-loop finiteness now follows almost immediately since only W survives on-shell and no invariants exist to this order.

It is amusing to note a point that has so far gone unnoticed in the literature. In ordinary gravity there are 14 algebraically independent scalars that can be formed from the Riemann tensor, of which 14 - 10 = 4 survive on-shell [e.g. Weinberg (1972)]. In supergravity there are just half as many! Seven may be formed from R, G, and W, of which 7 - 5 = 2 survive on-shell (Duff & Stelle, unpublished). They are

$$W_{\alpha\beta\gamma}W^{\alpha\beta\gamma} \quad \text{and} \quad \bar{W}_{\dot{\alpha}\dot{\beta}\dot{\gamma}}\bar{W}^{\dot{\alpha}\dot{\beta}\dot{\gamma}} ,$$

where $\bar{W}^{\dot{\alpha}\dot{\beta}\dot{\gamma}}$ is the complex conjugate of $W_{\alpha\beta\gamma}$. When integrated over superspace, these yield the two topological invariants $\chi \pm iP/2$ where $\chi$ is the Euler number, and P the Pontryagin number (see Section 4).

Algebraic independence, of course, is not sufficient for enumerating possible counterterms. New counterterms may be formed by taking products. For example (Ferrara & Zumino 1978 a)

$$W_{\alpha\beta\gamma}W^{\alpha\beta\gamma}\bar{W}_{\dot{\alpha}\dot{\beta}\dot{\gamma}}\bar{W}^{\dot{\alpha}\dot{\beta}\dot{\gamma}}$$

is just the three-loop (square of Bel-Robinson tensor) invariant discussed previously. Note, however, that since $W_{\alpha\beta\gamma}$ is anticommuting with three symmetric indices, $W^n = 0$ for $n > 4$ (Christensen et al. 1979). The real proliferation of possible counterterms occurs because new invariants can be formed from the supercovariant derivatives $D_\alpha$ and $\bar{D}_\alpha$. Their number can be considerably reduced, however, by appealing to another symmetry of our Lagrangian (2.6), namely $\gamma_5$ invariance. The remainder may then be classified, and one discovers in the process the interesting property that all vanish when the fields are self-dual or anti-self-dual, i.e. when either W or $\bar{W}$ vanishes (Christensen & Duff 1979; Christensen et al. 1979; Kallosh 1979 a,b). This is intimately connected with the phenomenon of helicity conservation in supergravity (Grisaru & Pendleton 1977; Christensen et al. 1979; Duff & Isham 1979, 1980; Duff 1979). Despite this reduction in number, non-vanishing invariants survive at every loop order $\geq 3$.

What about further matter couplings? Just as coupling random matter to pure gravity only made things worse, so coupling random numbers of N = 1 supermultiplets to N = 1 supergravity spoils even one loop finiteness (Fischler 1979; van Nieuwenhuizen & Vermaseren 1977). In ordinary gravity we learned that things improved only when the matter coupling was fixed by supersymmetry. Might the coupling of N = 1 supergravity to N = 1 supermatter lead to finiteness if this coupling is constrained by more supersymmetry? This leads us naturally to extended supergravity.

## 2.4 Extended supergravity

A detailed discussion of higher loop invariants which might act as counterterms in extended supergravity is given by Dr. Kallosh in this volume and here we confine ourselves to a few remarks. First we note that in spite of the extra supersymmetry, and in spite of the generalization of $\gamma_5$ invariance to combined $\gamma_5$ and duality invariance, higher loop invariants still exist which might act as counterterms (Deser & Kay 1978; Howe & Lindstrom 1981; Kallosh 1981; Howe et al. 1981; Stelle 1981 a). This is true, moreover, even for $N > 4$ (for which no matter multiplets exist) and including N = 8 which naturally stands out as the favourite candidate for finiteness. This seems a good time to pause, and question some of the assumptions which have underlied this programme so far.

Before looking for ways of reducing, or completely eliminating counterterms, let us first look on the negative side and ask whether we have not been too slick in simply listing on-shell invariants. (We recall first of all that we have ignored topological effects, ignored cosmological additions in our Lagrangian, ignored also the effects of boundary terms, and ignored spontaneously broken versions of supergravity; more of all this later.) To begin with, we have assumed that our regularization scheme preserves supersymmetry. Is this justified? Although conventional dimensional regularization is known not to, "dimensional reduction" seems better in this respect. See, for example, Siegel et al. (1981). However, it has recently been argued that dimensional reduction can be made consistent only at the expense of losing manifest supersymmetry at high loop order, for example at eight loops in N = 4 Yang-Mills (Avdeev et al. 1981). It may be that supersymmetry is in fact preserved but this issue has not yet been completely resolved.

Given that we have a good regularization scheme (see also Slavnov 1981) are we now justified in looking only at on-shell invariants as possible counterterms? This presupposes a background field method which manifestly preserved the supersymmetry, which in turn requires supersymmetric gauge-fixing and ghosts. Such a scheme must, in principle, exist even if we do not resort to it in actual loop calculations. Some of the problems involved have been discussed by De Wit and Grisaru (1979). Fortunately such a scheme does exist, at least for N = 1 supergravity where the superfield formalism (which incorporates the auxiliary fields) is completely known. The most comprehensive reference is Grisaru & Siegel (1981 a), and the techniques described there will probably turn out to be most efficient in practice as well as in principle. However, this brings us back to the question we have postponed so far: that of the auxiliary fields. Already for simple supergravity there is an *embarras de richesse* of different auxiliary field formulations (Breitenlohner 1977; Stelle & West 1978; Ferrara & van Nieuwenhuizen 1978 a; Sohnius & West 1981 b). Do they yield different quantum theories? They certainly yield different anomalies, as explained in Section 4, though there is a general belief that, anomalies aside, different auxiliary fields yield the same on-shell S-matrix elements. But what about extended supergravity with $N \geq 3$, and Yang-Mills with N = 4 where the auxiliary field structure is not only unknown but for which there is not even an existence proof? (See, for example, Taylor 1981 a,b,c.) Is it fair to list on-shell invariants as possible counterterms when their off-shell extension might not exist? The need for a thorough understanding of the auxiliary field structure in extended theories as a prerequisite for understanding the ultraviolet problem has been stressed by several authors. This said, however, it seems unlikely that the beautiful cancellations discussed in Section 3 would suddenly disappear when the auxiliary fields are included! Few would dispute, of course, that a complete auxiliary-field/superspace approach would make the task of investigating finiteness a whole lot easier.

In summary, it seems that listing on shell invariants as possible counterterms is probably correct as far as it goes. An entirely different question, of course, is whether it is sufficient. Turning now to a more positive approach we ask "Might the numerical coefficients of such counterterms turn out to be zero?".

## 3 SPIN SUM RULES AND VANISHING β-FUNCTIONS

### 3.1 Spin sum rules

In this section we concentrate on infinity cancellations known to occur in supersymmetric theories and supergravity which are not explicable by the "absence-of-invariant-as-counterterm arguments" discussed so far, i.e. the so-called "miraculous" cancellations whereby an invariant counterterm might exist but nevertheless appears with zero coefficient.

The most startling example of this phenomenon in supergravity is the vainishing of the one loop β function in gauged $N > 4$ extended theories (Christensen et al. 1980). At the time this result was published, the reason for this "miraculous" cancellation was unknown, but one was immediately reminded of the equally "miraculous" vanishing of the β-function in $N > 2$ supersymmetric Yang-Mills theories which is now known to hold to at least three loops (Avdeev et al. 1980; Grisaru et al. 1980; Caswell & Zanon 1980). Indeed, at the time, the complete Lagrangians for the gauged $N > 4$ theories had not yet been constructed. There was even doubt in some quarters whether gauged $N > 4$ supergravities actually existed; doubt which has now been dispelled by De Wit & Nicolai (1981 a,b).

The question now arises "What is so special about higher N values?" Since we do not yet have a complete superfield description of extended theories, let us examine the spin content in components. See Tables 1 and 2. As is well known, all supermultiplets share the property of having an equal number of Bose and Fermi degrees of freedom, i.e.

$$\sum_{\lambda} (-1)^{2\lambda}\, d(\lambda) = 0 \qquad (3.1)$$

where $\lambda$ is the helicity, ranging over 1, ½, 0, -½, -1 in Yang-Mills and 2, 3/2, 1, ½, 0, -½, -1, -3/2, -2 in supergravity, and $d(\lambda)$ is the number of states with helicity $\lambda$ in a supermultiplet. More remarkable, however, are the following generalizations, first published by Curtwright (1981):

$$\sum_{\lambda} (-1)^{2\lambda}\, d(\lambda)\, \lambda^{k} = 0\ , \quad N > k\ . \qquad (3.2)$$

These rules were first made known to me in 1980 by A. D'Adda, who conjectured that they were related to the vanishing β function. The Tables 1 and 2: Number of states $d(\lambda)$ with helicity $\lambda$ in supersymmetric Yang-Mills (1) and extended supergravity (2). The CPT conjugate multiplets must also be counted except for the self-conjugate N = 4 Yang-Mills and N = 8 supergravity. The numbers in brackets denote the quadratic Casimir invariant $C(\lambda)$.

Table 1

| $\lambda$ \ N | 1 | 2 | 3 | 4 |
|---|---|---|---|---|
| 1 | 1(c) | 1(c) | 1(c) | 1(c) |
| $\frac{1}{2}$ | 1(c) | 2(c) | 3(c) | 4(c) |
| 0 | | 1(c) | 3(c) | 6(c) |
| $-\frac{1}{2}$ | | | 1(c) | 4(c) |
| -1 | | | | 1(c) |

Table 2

| $\lambda$ \ N | 1 | 2 | 3 | 4 | 5 | 6 | 7 | 8 |
|---|---|---|---|---|---|---|---|---|
| 2 | 1(0) | 1(0) | 1(0) | 1(0) | 1(0) | 1(0) | 1(0) | 1(0) |
| $\frac{3}{2}$ | 1(0) | 2(1) | 3(1) | 4(1) | 5(1) | 6(1) | 7(1) | 8(1) |
| 1 | | 1(0) | 3(1) | 6(2) | 10(3) | 15(4) | 21(5) | 28(6) |
| $\frac{1}{2}$ | | | 1(0) | 4(1) | 10(3) | 20(6) | 35(10) | 56(15) |
| 0 | | | | 1(0) | 5(1) | 15(4) | 35(10) | 70(20) |
| $-\frac{1}{2}$ | | | | | 1(0) | 6(1) | 21(5) | 56(15) |
| -1 | | | | | | 1(0) | 7(1) | 28(6) |
| $-\frac{3}{2}$ | | | | | | | 1(0) | 8(1) |
| -2 | | | | | | | | 1(0) |

proof of this relation was carried out by Duff and Gibbons (unpublished) both for Yang-Mills and supergravity, independently of Curtwright. Since Curtwright's explanation of the supergravity result differs somewhat from ours, we shall here present both explanations and demonstrate their equivalence. There is one further difference, of a purely technical nature. Curtwright bases his $\beta$ function calculations on a Feynman graph analysis due to Hughes (1980). We shall instead adopt the arbitrary-spin background-field formalism of Christensen and Duff (1978 a, 1979, 1980).

### 3.2 Background field method

Consider a general field theory with fields denoted by the generic symbol $\hat{\phi}^i(x)$ and action denoted by $S[\hat{\phi}]$. If we make the background field split

$$\hat{\phi}(x) = \phi(x) + h(x) \tag{3.3}$$

and Taylor expand the action about the background field $\phi(x)$, we obtain

$$S[\phi + h] = S[\phi] + \int d^4x \left.\frac{\delta S}{\delta\hat{\phi}^i(x)}\right|_{\hat{\phi}=\phi} h^i(x) + \int d^4x\, d^4y\, h^i(x) \left.\frac{\delta^2 S}{\delta\hat{\phi}^i(x)\delta\hat{\phi}^j(y)}\right|_{\hat{\phi}=\phi} h^j(y) + O(h^3) . \tag{3.4}$$

Note that terms linear in the quantum field $h(x)$ are absent when $\phi$ is chosen to be a solution of the classical background field equations

$$\left.\frac{\delta S}{\delta\hat{\phi}}\right|_{\hat{\phi}=\phi} = 0 .$$

This will be important when we consider quantum gravity with a cosmological constant, as we must do in order to analyse gauged supergravity. To one loop order we need retain in Eq. (3.4) only these terms quadratic in h.

The quantum effective action $\Gamma[\phi]$ is now given by

$$e^{i\Gamma[\phi]} = e^{iS[\phi]} e^{iW[\phi]} , \tag{3.5}$$

where W is given by the functional integral

$$e^{iW[\phi]} = \int dh \exp (i \int d^4x \, \tfrac{1}{2} h^i \, \Delta_{ij} h^j) . \tag{3.6}$$

If h is a boson,

$$e^{iW[\phi]} = (\det \Delta)^{-1/2} . \tag{3.7}$$

$\Delta_{ij}$ is a second-order differentail operator determined by the second functional derivative of S in Eq. (3.4) and is a functional of the background field $\phi$. Graphically, the effective action $W[\phi]$ describes all one-loop graphs with closed h loops and external $\phi$ lines. Setting $\phi$ to be a solution of the classical field equations corresponds to going on-shell. It turns out that for the theories in which we are interested one can always arrange (e.g. by gauge fixing) that $\Delta_{ij}$ takes the simple form

$$\Delta^{ij} = -\delta^{ij} \nabla^\mu \nabla_\mu + X^{ij} , \tag{3.8}$$

where

$$\nabla_\mu h^i = \partial_\mu h^i + N_\mu{}^{ij} h_j \tag{3.9}$$

and where the matrices $X^{ij}$ and $N^{ij}$ are functionals of $\phi$ such that

$$N_\mu^{ij} = -N_\mu^{ji} , \quad X^{ij} = X^{ji} . \tag{3.10}$$

If, on the other hand, h is a fermion, we have

$$e^{iW[\phi]} = \det \not{D} = (\det \Delta)^{1/2} , \tag{3.11}$$

where

$$\Delta = \not{D}^2 = -\nabla^{\mu}\nabla_{\mu} + X \ , \tag{3.12}$$

and

$$\nabla_{\mu} = \partial_{\mu} + N_{\mu} \tag{3.13}$$

where $\not{D}$ is the Dirac operator in the $\phi$ background. For most purposes $\phi$ will be bosonic. Thus the fermion calculation can be converted into the same as the boson. In both cases we study operators of the form (3.8), but there is a change in sign in going from Eq. (3.7) to Eq. (3.11). [The exponent $\frac{1}{2}$ in Eq. (3.11) is replaced by $\frac{1}{4}$ if the spinor is real, i.e. Majorana.]

We are now in a position to write down, without proof, the one-loop counterterm $\mathcal{L}_{(1)}$ which must be added to render W finite. Defining the matrix

$$Y_{\mu\nu}{}^{ij} = -Y_{\mu\nu}{}^{ji} \tag{3.14}$$

by

$$\left[\nabla_{\mu}, \nabla_{\nu}\right] h^{i} = Y_{\mu\nu}{}^{ij} \, h_{j} \ , \tag{3.15}$$

i.e.

$$Y_{\mu\nu} = \partial_{\mu}N_{\nu} - \partial_{\nu}N_{\mu} + \left[N_{\mu}, N_{\nu}\right] , \tag{3.16}$$

then

$$\mathcal{L}_{(1)} = \pm \frac{1}{\varepsilon} \frac{1}{180(4\pi)^2} \sqrt{g} \ \mathrm{Tr} \left[ \mathbb{1}\left(R_{\mu\nu\rho\sigma}R^{\mu\nu\rho\sigma} - R_{\mu\nu}R^{\mu\nu} + \frac{5}{2} R^2\right) - 30RX + 90X^2 + 15Y_{\mu\nu}Y^{\mu\nu} \right] , \tag{3.17}$$

where $\pm$ refers to boson or fermion, respectively. In Eq. (3.17) we have allowed for the possibility of a non-vanishing gravitational background field in addition to any other background fields. This result can be obtained either by writing down the most general counterterm allowed and then fixing the numerical coefficients using momentum-space Feynman graphs or better (especially for the cosmological, topological and boundary terms) by the coordinate space "heat-kernel" expansion. A list of references may be found in Christensen and Duff (1979).

From Eq. (3.17), we see that the problem of computing one-loop counterterms is reduced to the problem of determining the matrices X and $Y_{\mu\nu}$ for the system in question. In the case where the gauge group is the internal Yang-Mills group the quantum field $h^i$ transforms like

$$h^i \to \Omega^{ij} h_j \ , \tag{3.18}$$

where

$$\Omega^{ij} = \exp \varepsilon^a(x)\, T_a{}^{ij} \tag{3.19}$$

and where the generators $T_a$ obey

$$\left[T_a, T_b\right] = f_{ab}{}^c T_c \ . \tag{3.20}$$

The covariant derivative is

$$\nabla_\mu h^i = \partial_\mu h^i + A_\mu{}^a T_a{}^{ij} h_j \tag{3.21}$$

and

$$Y_{\mu\nu} = \partial_\mu A_\nu - \partial_\nu A_\mu + \left[A_\mu, A_\nu\right] \tag{3.22}$$

$$= F_{\mu\nu}{}^a T_a \ , \tag{3.23}$$

with $F_{\mu\nu}{}^a$ the Yang-Mills field strength. In the case where the gauge group is the external Lorentz group, we have

$$h^\alpha \to \Omega^{\alpha\beta} h_\beta \ , \tag{3.24}$$

where $\alpha$ is an arbitrary spin index, and where

$$\Omega^{\alpha\beta} = \exp \omega^{ab}(x)\ \Sigma_{ab}{}^{\alpha\beta} . \tag{3.25}$$

The Lorentz generators $\Sigma_{ab}$ obey

$$\left[\Sigma_{ab}, \Sigma_{cd}\right] = \frac{1}{2}\left(\eta_{cb}\Sigma_{ad} - \eta_{ca}\Sigma_{bd} + \eta_{bd}\Sigma_{ca} - \eta_{da}\Sigma_{cb}\right) \tag{3.26}$$

with

$$\Sigma_{ab} = - \Sigma_{ba} . \tag{3.27}$$

The covariant derivative is now

$$\nabla_\mu h^\alpha = \partial_\mu h^\alpha + \omega_\mu{}^{ab}\Sigma_{ab}{}^{\alpha\beta}h_\beta \tag{3.28}$$

and

$$Y_{\mu\nu} = \partial_\mu\omega_\nu - \partial_\nu\omega_\mu + \left[\omega_\mu, \omega_\nu\right] \tag{3.29}$$

$$= R_{\mu\nu}{}^{ab}\Sigma_{ab} \tag{3.30}$$

with $R_{\mu\nu}{}^{ab}$ the Riemann tensor.

In Yang-Mills theories $Y_{\mu\nu}$ depends on the internal representation of whatever quantum field if going around the loop as in Eq. (3.23) but is independent of the spin. The spin dependence enters via

$$X = \Sigma^{\mu\nu}F_{\mu\nu}{}^{a}T_a . \tag{3.31}$$

Remarkably, this formula is valid whatever the spin, i.e. $h^i$ could be spin-0, spin-½, or else the spin-1 gauge field itself (or its spin-0 ghosts). Similarly in gravity $Y_{\mu\nu}$ is already linear in $\Sigma_{\mu\nu}$ as in Eq. (3.30), whereas X is quadratic, typically

$$X = \Sigma^{\mu\nu}R_{\mu\nu}{}^{ab}\Sigma_{ab} . \tag{3.32}$$

It remains to evaluate the Tr operation in Eq. (3.17). These arbitrary spin formulae and the evaluation of the necessary traces may be found in Christensen and Duff (1979) and we shall now summarize the results.

### 3.3 Arbitrary spin formalism

Irreducible representations of the Lorentz group are labelled (A,B) where the non-negative numbers A and B take on integer or half-integer values. The dimensionality of the representation, or number of degrees of freedom, is

$$D(A,B) = (2A+1)(2B+1) \tag{3.31}$$

$$= (s+1)^2 - t^2 \; , \tag{3.32}$$

where the spin s is given by

$$s \equiv A + B \tag{3.33}$$

and where

$$t \equiv A - B \; , \quad -s \leq t \leq s \; . \tag{3.34}$$

However, the particles which appear in Tables 1 and 2 are not in general described by a single irreducible representation. They each correspond to two degrees of freedom (counting scalars as complex) and for $s \geq 1$ it is necessary to include Fadeev-Popov ghost subtractions in order to arrive at just two helicity states. The general rule (Duff 1979) is first to compute the contributions to Eq. (3.17) from a representation (A,B) add to it the contribution from (A-1, B-1) and then subtract twice the contribution from $(A-\frac{1}{2}, B-\frac{1}{2})$. For example, the correct degree-of-freedom count is given not by

$$\mathrm{Tr}\, \mathbb{1} = D(A,B) = (2A+1)(2B+1) \tag{3.35}$$

but rather by

$$D'(A,B) = D(A,B) + D(A-1, B-1) - 2D(A-\tfrac{1}{2}, B-\tfrac{1}{2})$$

$$= 2 \; . \tag{3.36}$$

In tracing products of $\Sigma$'s, one also encounters the functions

$$E_{\pm}(A,B) = D[A(A+1) \pm B(B+1)] \tag{3.37}$$

from two $\Sigma$'s, and

$$F_{\pm}(A,B) = D[A(A+1)(2A-1)(2A+3) \pm B(B+1)(2B-1)(2B+3)] \tag{3.38}$$

$$G(A,B) = D[A(A+1) + B(B+1)]^2 \tag{3.39}$$

from four $\Sigma$'s (which is as many as we ever need, at least to this one-loop order). If we translate these into functions of s and t and then calculate the corresponding primed quantities as in Eq. (3.36), one obtains

$$D' = 2$$

$$E'_{+} = 6s^2$$

$$F'_{+} = -15s^2 + 15s^4 + t^2\,(5 - 5t^2 + 30s^2)$$

$$E'_{-} = 6st$$

$$F'_{-} = -10st + 40s^3t$$

$$G' = \frac{3}{2}\,s^2 + \frac{15}{2}\,s^4 + t^2\left(\frac{1}{2} - \frac{t^2}{2} + 3s^2\right) . \tag{3.40}$$

As we shall see, both in this section and in Section 4, these functions are all we need to write down the $\beta$-functions, the one-loop counterterms, the conformal anomalies, and the axial anomalies for fields of arbitrary spin, in both Yang-Mills and gravity.

### 3.4 Supersymmetric Yang-Mills: $\beta = 0$ for $N > 2$

In flat space, the one-loop counterterm Eq. (3.17) reduces to

$$\mathcal{L}_{(1)} = \pm\,\frac{1}{\varepsilon}\,\frac{1}{(4\pi)^2}\,\mathrm{Tr}\left[\frac{1}{2}\,X^2 + \frac{1}{12}\,Y_{\mu\nu}Y^{\mu\nu}\right] . \tag{3.41}$$

For supersymmetric Yang-Mills, the particles going around the loop will be the spin-1 quantum gauge fields, their spin-0 ghosts, spin-½ fermions, and physical spin-0 scalars. In each case X and $Y_{\mu\nu}$ are given by Eq. (3.23) and Eq. (3.31). Hence

$$\mathrm{Tr}\ Y_{\mu\nu}Y^{\mu\nu} = -D\ F_{\mu\nu}{}^{a}F^{\mu\nu a}C \tag{3.42}$$

$$\mathrm{Tr}\ X^2 = \frac{2}{3}\left(E_+F_{\mu\nu}{}^{a}F^{\mu\nu a} + E_-{}^{*}F_{\mu\nu}{}^{a}F^{\mu\nu a}\right)C\ , \tag{3.43}$$

where $F_{\mu\nu}{}^{a}$ is the background field strength and ${}^{*}F_{\mu\nu}{}^{a}$ is its dual

$${}^{*}F_{\mu\nu}{}^{a} \equiv \frac{1}{2}\,\varepsilon_{\mu\nu\rho\sigma}F^{\rho\sigma a}\ , \tag{3.44}$$

where D and $E_\pm$ are given by Eqs. (3.35) and (3.37), and where C is the second-order Casimir

$$-\delta^{ab}C = \mathrm{Tr}\ T^{a}T^{b}\ . \tag{3.45}$$

For physical fields we replace D and $E_\pm$ by $D'$ and $E'_\pm$ of Eq. (3.40) to obtain the counterterm

$$\mathcal{L}_{(1)} = \frac{1}{\varepsilon}\,\frac{(-1)^{2s}}{(4\pi)^2}\,e^2\left[\left(-\frac{1}{6} + 2s^2\right)F_{\mu\nu}{}^{a}F^{\mu\nu a} + 2st\,{}^{*}F_{\mu\nu}{}^{a}F^{\mu\nu a}\right]C\ , \tag{3.46}$$

where we have rescaled $A_\mu \to eA_\mu$ to introduce the gauge coupling constant e. The topological counterterm

$$P = \frac{e^2}{16\pi^2}\int d^4x\ {}^{*}F_{\mu\nu}{}^{a}F^{\mu\nu a} \tag{3.47}$$

need not concern us yet and we turn instead to the one-loop $\beta(e)$ function. The contribution to $\beta(e)$ from a particle of spin s is given by the coefficient of $-\varepsilon^{-1}e^{-1}F_{\mu\nu}{}^{a}F^{\mu\nu a}$ in Eq. (3.46), i.e.

$$\beta(s) = \frac{e^3}{96\pi^2}\,(-1)^{2s}(1 - 12s^2)C\ . \tag{3.48}$$

This is the same result as that of Hughes (1980) and Curtwright (1981), who interpret the first and second terms in Eq. (3.48) as the "convective charge" and "magnetic moment" contributions, respectively. The asymptotic freedom of pure Yang-Mills (i.e. $\beta < 0$ for $s = 1$) then follows immediately as a consequence of the negative magnetic moment term dominating the positive convective charge term; an interpretation which had, in fact, already been anticipated by Salam and Strathdee (1975). If there are several fields in the theory, the complete one-loop $\beta$-function is obtained by multiplying Eq. (3.48) by d(s), the number of fields of spin s, and summing over spins

$$\beta = \sum_s d(s)\beta(s) \ . \tag{3.49}$$

In the case of supersymmetric Yang-Mills, the internal symmetry factor C is the same for all spins since each belongs to the same (adjoint) representation of the gauge group. Now, of course, we may invoke the spin sum rules [Eq. (3.2)], in particular

$$\sum_s (-1)^{2s} d(s) = 0 \ , \quad \forall N \tag{3.50}$$

$$\sum_s (-1)^{2s} d(s) s^2 = 0 \ , \quad N > 2 \ . \tag{3.51}$$

(Note that for sum rules with even powers of $\lambda$ we may replace the helicity $\lambda$ by the spin s provided we include the CPT conjugates and sum over s = 0, ½, and 1 with the understanding that d(0) counts the number of complex scalars.)

The crucial observation is that the vanishing of the one-loop $\beta$-function for N = 4 is no longer miraculous but an obvious consequence of the sum rules [Eqs. (3.50) and (3.51)] applied to the arbitrary spin results [Eqs. (3.48) and (3.49)]. Note that the first term in Eq. (3.48) cancels for all N, whereas the second term cancels only for N > 2. (For N = 1 and N = 2 this second term demonstrates asymptotic freedom.)

What about higher loops? Sum rules apart, we know that the N = 4 theory is finite to at least three loops and arguments have been put forward to suggest that this is true to all orders (Ferrara & Zumino 1978 b; Sohnius & West 1981 a). Can we therefore give a sum-rule explanation? First we note that beyond one loop, it no longer makes sense to attribute a contribution to $\beta$ from each individual spin as in Eq. (3.48), since the contributions from different spins will mix. So we would expect to have to sum over spins as in Eq. (3.49). Thus one might make an all-orders guess of

$$\beta = \sum_{\lambda} (-1)^{2\lambda} d(\lambda)\left[a + b\lambda^2\right] , \tag{3.52}$$

with a and b universal functions of the coupling constant, since this reduces to the correct result at one loop with

$$a = \frac{e^3}{96\pi^2} C , \quad b = -\frac{e^3}{6\pi^2} C . \tag{3.53}$$

We have included no powers of $\lambda$ greater than two under the summation; otherwise the sum rules [Eqs. (3.50) and (3.51)] could not explain the vanishing $\beta$ at 2 and 3 loops in N = 4. Secondly, we would not expect such a formula to hold for non-supersymmetric theories beyond one loop, because the gauge, Yukawa, and quartic scalar coupling constants (which coincide for supersymmetric Yang-Mills theories) are in general different. In this supersymmetric case, moreover, the a term now vanishes for all N by virtue of Eq. (3.50). Having said all this, however, the resulting formula

$$\beta = b \sum_{\lambda} (-1)^{2\lambda} \lambda^2 ,$$

though consistent with the vanishing $\beta$ for N = 4, fails to account for another curious result, namely the vanishing of the two-loop contribution to $\beta$ for N = 2 but not N = 1 (Jones 1980). To explain this, we would have to give up the universality of b and instead introduce another factor $\Sigma_\lambda (-1)^{2\lambda}\lambda$ into its two-loop contribution which from Eq. (3.2) would vanish for N = 2 but not N = 1. Thus it seems that we must abandon guess-work about higher loop order for the time being since the simplest guess Eq. (3.52) does not seem to work. Let us instead remain at one loop and examine supergravity.

### 3.5 Gauged extended supergravity: $\beta = 0$ for $N > 4$

Extended supergravities with N gravitini exhibit a global SO(N) symmetry. Moreover, the N(N-1)/2 spin-1 fields lie in the adjoint representation, see Table 2. (For N = 6, there is one extra vector.) This suggests a possible gauging of this SO(N) symmetry whereby ordinary derivatives are replaced by SO(N) Yang-Mills covariant derivatives with a corresponding covariantization of the spin-1 kinetic term. In this way we acquire a new dimensionless coupling constant e in addition to the dimensionful coupling constant $\kappa$ already present. It remains to show, however, that one can make other e-dependent additions to the Lagrangian in such a way as to maintain (with e-dependent corrections to the transformation laws) the N-fold supersymmetry. That this can indeed be carried out in a consistent fashion was demonstrated by Freedman & Das (1977) and Fradkin & Vasiliev (1976) for N = 2 and N = 3; by Das et al. (1977) and Freedman & Schwarz (1978) for N = 4; and most recently by De Wit & Nicolai (1981 a,b) for N = 5, 6, 7, and 8. The graded Poincaré algebra gets replaced by the graded de Sitter algebra, and the Lagrangian acquires a cosmological constant $\Lambda$ given by

$$\Lambda = -\frac{6e^2}{\kappa^2} \tag{3.54}$$

and gravitino "mass" term given by

$$m^2 = \frac{2e^2}{\kappa^2} . \tag{3.55}$$

Equation (3.54) has an interesting consequence. By linking the gauge coupling constant to the cosmological constant, the renormalization of $\kappa^2\Lambda$ determines the $\beta(e)$-function. Thus to calculate $\beta$ one may proceed in one of two ways. Either (a) compute the usual charge renormalization effects, i.e. the coefficient of the Yang-Mills Tr $\sqrt{g}\, F_{\mu\nu}F^{\mu\nu}$ counterterm which receives contributions from spins 0, $\frac{1}{2}$, 1, and $\frac{3}{2}$, but not 2, since the graviton is a singlet, or (b) compute the cosmological renormalization, i.e. the coefficient of the $\sqrt{g}$ counterterm, which receives contributions from all spins including gravity. By supersymmetry, one determines the other.

Since the one-loop cosmological renormalization had already been carried out by Christensen and Duff (1980) for spins 0, ½, 1, and 2, it was natural to attempt the latter of these approaches first by extending these calculations to spin 3/2 and hence to gauged extended supergravity (Christensen et al. 1980). Remarkably, we found that while $\beta > 0$ for $N \leq 4$, $\beta = 0$ for $N > 4$! Subsequently Curtwright (1981) arrived at the same result using the first of these approaches.

We shall now describe both calculations, show how they lead to equivalent results, and in each case give a spin sum rule explanation for the vanishing $\beta$-function when $N > 4$. First, however, it is necessary to say a few words about quantizing gravity with a cosmological constant.

### 3.6 The cosmological constant

Although the renormalizability properties of quantum gravity and supergravity have received considerable attention over the last few years, almost all these investigations have confined their attention to theories with vanishing cosmological constant, $\Lambda$. Contrary to a popular school of thought, however, the calculation of the quantum effective action when $\Lambda \neq 0$, i.e. when the gravitational action is

$$S = -\frac{1}{2\kappa^2} \int d^4x \sqrt{g}\,(R - 2\Lambda) \ , \tag{3.56}$$

is no more difficult than when $\Lambda = 0$, provided one consistently expands about a background field satisfying the Einstein equation with a $\Lambda$ term

$$R_{\mu\nu} = \Lambda g_{\mu\nu} \ . \tag{3.57}$$

See Christensen and Duff (1980). [In particular one must avoid an expansion about flat space. This is not the correct ground-state when $\Lambda \neq 0$. Attempting the expansion $g_{\mu\nu} = \eta_{\mu\nu} + \kappa h_{\mu\nu}$ leads to problems both in the term linear in $h_{\mu\nu}$ and the term quadratic in $h_{\mu\nu}$ which appear in the expansion of a $\sqrt{g}$ Lagrangian. The former gives rise to awkward ill-defined tadpole graphs and the latter to massive ghosts.] As explained in Section 3.2, terms linear in the quantum field h are absent when the background field is a solution of the classical field equations [see Eq. (3.4)]. Terms quadratic in h, which govern the one-loop calculations, are then determined in a suitable gauge by the operators

$$\Delta h_{\mu\nu} = -\nabla^\rho \nabla_\rho h_{\mu\nu} - 2R_{\mu\rho\nu\sigma} h^{\rho\sigma} \tag{3.58}$$

in the case of the graviton and

$$\Delta h_\mu = -\nabla^\mu \nabla_\rho h_\mu - \Lambda h_\mu \tag{3.59}$$

in the case of the spin-1 Fadeev-Popov ghosts. The one-loop counterterms then follow from Eq. (3.17). They will be of the form $R_{\mu\nu\rho\sigma}R^{\mu\nu\rho\sigma}$, $R_{\mu\nu}R^{\mu\nu}$, $R^2$, $\Lambda R$, and $\Lambda^2$ with gauge-dependent coefficients. Gauge independence is achieved after going "on-shell" by use of Eq. (3.57), resulting in the one-loop counterterm

$$S_{(1)} = \frac{1}{\varepsilon}\left[A\chi - \frac{B\kappa^2\Lambda}{12\pi^2} S\right], \tag{3.60}$$

where A and B are numerical coefficients,

$$\chi \equiv \frac{1}{32\pi^2}\int d^4x \sqrt{g}\left(R_{\mu\nu\rho\sigma}R^{\mu\nu\rho\sigma} - 4R_{\mu\nu}R^{\mu\nu} + R^2\right) \tag{3.61}$$

is the Euler number (see Section 4), and S is the classical action on shell. Explicit calculation (Christensen & Duff 1980) yields

$$A = \frac{106}{45}, \quad B = -\frac{87}{10}. \tag{3.62}$$

Thus, in contrast to the case $\Lambda = 0$ discussed in Section 2, pure gravity with a $\Lambda$ term is no longer one-loop finite (in the non-topological sense) because $B \neq 0$.

One may now repeat the exercise for simple (N = 1) supergravity with a gravitino mass term

$$S = \int d^4x \det e_\mu{}^a \left[-\frac{1}{2\kappa^2} R + \frac{1}{2}\varepsilon^{\mu\nu\rho\sigma}\bar\psi_\mu \gamma_5 \gamma_\nu D_\rho \psi_\sigma \right.$$
$$\left. + m\,\bar\psi_\mu \sigma^{\mu\nu}\psi_\nu + \frac{1}{3}(S^2 + P^2 - b^\mu b_\mu) + \frac{2m}{\kappa} S\right]. \tag{3.63}$$

Elimination of the auxiliary field S yields a cosmological constant $\Lambda = -3m^2$. See, for example, van Nieuwenhuizen (1979). Previous discussions of supergravity with a cosmological constant may be found in MacDowell & Mansouri (1977), Deser & Zumino (1977), and Townsend (1977). The one-loop counterterm will again be of the form (3.60), where S is now given by Eq. (3.63) on shell. The coefficients A and B will now receive contributions from both the graviton and the gravitino (with its appropriate mass parameter). Explicit calculation (Christensen et al. 1980) yields

$$A = \frac{41}{24}, \quad B = -\frac{77}{12}, \tag{3.64}$$

and once again in contrast to the case $\Lambda = 0 = m$ discussed in Section 2, simple supergravity is no longer one-loop finite.

We may now combine these results with those for spins 1, ½, and 0 (Christensen & Duff 1980) and apply them to the extended SO(N) theories with gauged internal symmetry. The cosmological coefficient B now takes on a new significance: by supersymmetry it also determines the renormalization of the gauge coupling constant e. Combining Eqs. (3.54) and (3.60) we have

$$S_{(1)} = \frac{1}{\varepsilon}\left[A\chi + B\frac{e^2}{2\pi^2}S\right], \tag{3.65}$$

where the classical action S will now contain a spin-1 gauge field contribution $\mathrm{Tr}\, F_{\mu\nu}F^{\mu\nu}$. The signal for asymptotic freedom is $B > 0$.

Only the kinetic terms in the classical Lagrangian are needed to fix the contributions to A from fields of different spin. See Christensen and Duff (1978) and references therein. To calculate B we also require knowledge of the mass terms. All particles must be massless for all N if, as we are assuming, supersymmetry is not spontaneously broken. There is, however, an "apparent mass" parameter for the gravitinos given by Eq. (3.55). No such terms are present for the spin-1 or spin-½ fields but the spin-0 fields, which make their appearance for $N \geq 4$, require greater care. The spin-2, spin-0 coupling is known to be of the form

$$\mathcal{L} = -\frac{1}{2\kappa^2}\sqrt{g}\,(R-2\Lambda) + \frac{1}{2}\sqrt{g}\,\phi^i\left(-\Box + \frac{2\Lambda}{3}\right)\phi^i + O(\phi^3)\ , \quad (3.66)$$

i.e. minimal coupling with a mass term. (The range of the index i is given by Table 2.) However, one could equally well use

$$\mathcal{L} = -\frac{1}{2\kappa^2}\sqrt{g}\,(R-2\Lambda) + \frac{1}{2}\sqrt{g}\,\phi^i\left(-\Box + \frac{R}{6}\right)\phi^i + O(\phi^3)\ , \quad (3.67)$$

i.e. conformal coupling with no mass term. The equivalence is seen by making a Weyl rescaling in the Lagrangian (3.67) of the form

$$g_{\mu\nu} \to \Omega^2 g_{\mu\nu}\ , \quad \phi^i \to \Omega^{-1}\phi^i\ ; \quad \Omega^2 \equiv 1 + \frac{\kappa^2}{6}\phi^i\phi^i\ , \quad (3.68)$$

which yields the Lagrangian (3.66). Both Lagrangians yield the same B coefficient since the field equations imply $R = 4\Lambda + \ldots$ . We note in passing that the mass-term properties discussed above were, at the time the B coefficients were first calculated (Christensen et al. 1980), assumed for $N > 4$ on the basis of an extrapolation from the known $N = 4$ mass terms (Das et al. 1977; Freedman & Schwarz 1978). The existence of these gauged $N > 4$ theories together with their mass-term properties has subsequently been confirmed by De Wit & Nicolai (1981 a,b). [In none of the theories discussed here does the scalar potential contain terms linear in $\phi$. In this respect they differ from the alternative $N = 4$ theory of Freedman & Schwarz (1978), which has a chiral $SU(2) \times SU(2)$ symmetry with two independent coupling constants, and which we shall not discuss. However, all the gauged theories with scalars apparently suffer from a potential $V(\phi)$ which is not bounded below. But since $V(\phi)$ is intimately connected with $\Lambda$, it may be that the criterion for stability is different when $\Lambda \neq 0$. After all, we have already argued that an x-independent vacuum expectation value for $g_{\mu\nu}$ cannot provide the correct ground state when $\Lambda \neq 0$ and the same may well be true of the scalars $\phi^i(x)$. This is deserving of further study.]

The results of calculating the B coefficient for fields of spin s are summarized in Table 3. The combined results for extended supergravity then follow from the particle content of Table 2 and are listed in Table 4. The vanishing of B for $N > 4$ seems, at first sight, miraculous.

Table 3

Contributions to the cosmological B coefficient and charge renormalization β-function for fields of spin s. (For convenience, the spin-0 result is quoted for a two-component, i.e. complex, scalar)

| S | 60B | $96\pi^2\beta e^{-3}$ |
|---|---|---|
| 0 | -2 | C(0) |
| $\frac{1}{2}$ | -3 | 2 C($\frac{1}{2}$) |
| 1 | -12 | -11 C(1) |
| $\frac{3}{2}$ | 137 | 26 C($\frac{3}{2}$) |
| 2 | -522 | 0 |

Table 4

The values of B and β in extended supergravity, demonstrating the equivalence of the two calculational methods

| N | B | $-16\pi^2\beta e^{-3}$ |
|---|---|---|
| 1 | $-\frac{77}{12}$ | - |
| 2 | $-\frac{13}{3}$ | $-\frac{13}{3}$ |
| 3 | $-\frac{5}{2}$ | $-\frac{5}{2}$ |
| 4 | -1 | -1 |
| 5 | 0 | 0 |
| 6 | 0 | 0 |
| 7 | 0 | 0 |
| 8 | 0 | 0 |

However, we note that the B's of Table 3 are described by the quartic polynominal

$$60B(s) = (-1)^{2s}[-2 + 30s^2 - 40\, s^4] \tag{3.69}$$

$$= \frac{1}{3}\,(-3D' + 25E'_+ - 20G' + 2F'_+)\ , \tag{3.70}$$

where $D'$, $E'_+$, $G'$, and $F'_+$ are given by Eq. (3.40), as may be verified by an arbitrary spin background field calculation along the lines of that already described for the Yang-Mills field. The spin sum rules

$$\sum_s\ (-1)^{2s} d(s) s^k = 0\ , \quad N > k\ , \tag{3.71}$$

may now be invoked to explain the vanishing of B for $N > 4$. We note incidentally that we encounter a polynomial in spin of degree 4 in gravity in contrast to one of degree 2 in Yang-Mills. This is due to the extra Lorentz generators in the matrices X and Y of Eqs. (3.30) and (3.32) compared with those of Eqs. (3.23) and (3.31). A naïve argument for higher loops might be to note that, since Yang-Mills is renormalizable, one could never encounter more than $X^2 \sim s^2$ in the higher loop counterterm and that the sum rules could then keep $N > 2$ finite to all orders. The same naïve power counting argument applied to gravity, on the other hand, would yield $X^3 \sim s^6$ at two loops, $X^4 \sim s^8$ at three loops, etc. And even the $N = 8$ sum rule fails at $s^8$. However, we have already seen the danger of extrapolation to higher loops based on guess-work, and a deeper understanding of the infinity cancellations is still required.

### 3.7 Charge renormalization

Having derived the supergravity β function via the cosmological method, we now discuss the alternative method due to Curtwright (1981) who simply extrapolated the $\beta(s)$ of Eq. (3.48) to the case of spin $^3/_2$ (an extrapolation which can be justified by the arbitrary spin background field calculations). This yields the β coefficients of Table 3. As is well-known, the spin-0 and spin-$^1/_2$ contributions enter with the opposite sign to the spin-1 gauge field. However, we learn that the spin-$^3/_2$ field also enters with the opposite sign and works against asymptotic freedom.

When applied to extended supergravity multiplets we obtain the results of Table 4 in complete agreement with the cosmological method. What is not so obvious, however, is the spin sum rule explanation of vanishing $\beta$ for $N > 4$. To give such an explanation we need the new rules (Curtwright 1981)

$$\sum_{\lambda} (-1)^{2\lambda} d(\lambda)\lambda^{k} C(\lambda) = 0 , \quad N > k + 2 , \tag{3.72}$$

which, it can be shown, follow as a consequence of the old rules (3.2). Thus although the $\beta(s)$ of Eq. (3.48) is only quadratic in spin, the Casimir invariant is now spin dependent and we again find $\beta = 0$ for $N > 4$.

To understand a little better how two such apparently different approaches lead satisfactorily to the same result for all N, we note that in the cosmological method we are calculating the $\sqrt{g}$ counterterm, whereas in the charge renormalization method we are calculating the $\sqrt{g}\ \mathrm{Tr}\ F_{\mu\nu}F^{\mu\nu}$ counterterm but, by supersymmetry, they enter in a fixed ratio. In fact, apart from $\chi$, the only counterterm surviving on shell is the classical action itself [see Eq. (3.65)]. In the cosmological method it appears with coefficient $B\ e^2(2\pi^2\varepsilon)^{-1}$ and in the charge renormalization method with coefficient $-8\beta(e\varepsilon)^{-1}$. Hence

$$B = -\frac{16\pi^2}{e^3}\,\beta , \tag{3.73}$$

a result confirmed by the explicit calculations summarized in Table 4. Remember, moreover, that graviton loops were used to compute the left-hand side of this equation but not the right-hand side!

There remains one possible question mark concerning the validity of the charge renormalization method and the extrapolation of the Yang-Mills $\beta(s)$ of Eq. (3.48) to the case of curved space theories with a cosmological constant. The $\beta$-function, i.e. that which is given by the coefficient of the $e^2\ \mathrm{Tr}\ F_{\mu\nu}F^{\mu\nu}$ counterterm, in principle receives contributions from two different sources. In addition to the usual charge renormalization effects, there might alse be a one-loop counterterm of the form $\kappa^2 R\ \mathrm{Tr}\ F_{\mu\nu}F^{\mu\nu}$. By using the field equations $R = 4\Lambda + \ldots$ , together with $\kappa^2\Lambda = -6e^2$ this is converted into an extra $e^2\ \mathrm{Tr}\ F_{\mu\nu}F^{\mu\nu}$ term. The fact that agreement with the cosmological method has been achieved without taking this into account means that, in one-loop supergravity at least, such counterterms must be absent. In fact, all Einstein-Yang-Mills theories avoid this problem at one loop order as can be seen by taking the $e \to 0$ limit and exploiting the Einstein-Maxwell duality arguments (Deser 1981). However, the $e \to 0$ argument cannot be invoked at higher loops where counterterms involving higher powers of the curvature with e dependent coefficients might yield $F^{\mu\nu}F_{\mu\nu}$ on shell. These might lead to a bizarre situation where flat space theories change their β function when coupled to gravity with a cosmological constant. One suspects that such terms are always absent in supergravity.

It is sometimes said that gravity theories, if they are to make sense, cannot be renormalizable but can only be finite owing to the dimensionful coupling constant $\kappa$. This ceases to be true, of course, when one allows for another dimensionless coupling constant e. Indeed the $N \leq 4$ theories discussed here are one-loop renormalizable in the sense that the non-vanishing counterterm is proportional to the classical action. This will continue to hold at two loops since the ungauged ($e = 0$) theories are known to be two-loop finite. Of course, the $N > 4$ theories stand a chance of being completely finite even after gauging.

Finally, although we have applied the spin sum rules (3.2) and (3.72) to the known Lagrangian field theories of supersymmetric Yang-Mills and gauged SO(N) supergravity, we should point out that from the purely group theoretical point of view, they have a much wider validity. As discussed by Curtwright (1981), they apply to all supermultiplets with either SO(N) or SU(N) internal symmetry, for example to the massive $N = 8$ supercurrent multiplet (Ellis et al. 1980; Deredinger et al. 1981), which features in the bound-state interpretation of presently observed quarks, leptons, and bosons.

### 3.8 Mass sum rules and symmetry breaking

The spin sum rules discussed above bear a remarkable resemblance to the mass sum rules which were already known in supergravity spontaneously broken via dimensional reduction. Consider the ungauged extended supergravities spontaneously broken through reduction from five dimensions (Cremmer et al. 1979; Ferrara & Zumino 1979). One has

$$\sum_{s} (-1)^{2s}(2s+1)M_s^{2k} = 0 \ , \quad N > 2k \ , \tag{3.74}$$

where $M_s$ is the mass of the particle with spin s.

In contrast to the (unbroken) gauged supergravities, these theories have zero $\Lambda$ at the tree level but will acquire one through radiative corrections. Owing to the mass sum rules, however, the induced cosmological constant is finite for $N \geq 6$. For further applications to spontaneously broken N = 8, see the lectures by van Nieuwenhuizen (1981 b).

## 4 ANOMALIES AND CHIRAL SUPERFIELDS

### 4.1 Axial and conformal anomalies for arbitrary spin

The arbitrary spin background field formalism discussed in Section 3 may also be used to determine the one-loop anomalies in ordinary (i.e. ungauged) supergravity. Consider, for example, the conformal anomaly in the trace of the stress tensor (Capper & Duff 1974 b; Deser et al. 1976; Duff 1977)

$$T^{\mu}{}_{\mu} = A\,\frac{1}{32\pi^2}\,{}^{*}R^{*}_{\mu\nu\rho\sigma}R^{\mu\nu\rho\sigma} + a\,\frac{e^2}{16\pi^2}\,F^{*}_{\mu\nu}{}^{a}F^{\mu\nu}{}_{a}C\ , \qquad (4.1)$$

where A is the same coefficient that appears in the topological counterterms (3.60) and (3.65) and a determines the Yang-Mills counterterm (3.46). One can show (Chistensen & Duff 1979) that

$$\begin{aligned} 12a &= (-1)^{2s}[-D' + 4E'_{+}] \\ 360A &= (-1)^{2s}[4D' - 10E'_{+} + 6F'_{+}]\ , \end{aligned} \qquad (4.2)$$

where $D'$, $E'_{+}$, and $F'_{+}$ are given by Eq. (3.40). These yield

$$\begin{aligned} 6a &= (-1)^{2s}[-1 + 12s^2] \\ 360A &= (-1)^{2s}[8 - 150s^2 + 90s^4 + 30t^2(1 - t^2 + 6s^2)]\ . \end{aligned} \qquad (4.3)$$

This A coefficient corresponds to the entries $e_{\mu}{}^{a}$, $\psi_{\mu}$, $\phi_{\mu}$, $\chi$, and $\phi$ in the first column of Table 5. [Note, incidentally, that if we drop the $F'_{+}$ term in Eq. (4.2), we obtain the simpler expression

$$360A' = (-1)^{2s}(8 - 60s^2)\ , \qquad (4.4)$$

which from the sum rules of Section 4 would yield a vanishing result for supermultiplets with $N > 2$. The corresponding numbers are given in the second column of Table 5. Note that $\Delta A = A - A' =$ integer. We shall return to the significance of this shortly.]

Table 5

Contributions to the A and A′ coefficients from the component fields of supergravity. The numbers correspond to those of physical fields (i.e. after gauge-fixing and after ghost subtractions).

| | 360A | 360A′ | ΔA |
|---|---|---|---|
| $e_{\mu a}$ | 848 | -232 | 3 |
| $\psi_\mu$ | -233 | 127 | -1 |
| $\phi_\mu$ | - 52 | - 52 | 0 |
| $\chi$ | 7 | 7 | 0 |
| $\phi$ | 4 | 4 | 0 |
| $\phi_{\mu\nu}$ | 364 | 4 | 1 |
| $\phi_{\mu\nu\rho}$ | -720 | 0 | -2 |

Similarly, the axial anomalies are given by the same combinations of $E'_-$ in the case of Yang-Mills and $E'_-$ and $F'_-$ in the case of gravity

$$\nabla^\mu J^5_\mu = \left(\frac{-E'_-}{6} + \frac{F'_-}{10}\right) \frac{1}{48\pi^2} R^*{}_{\mu\nu\rho\sigma} R^{\mu\nu\rho\sigma} + \frac{2E'_-}{3} \frac{e^2}{8\pi^2} \mathrm{Tr}\, F^*{}_{\mu\nu}{}^{\mu\nu} \ . \quad (4.5)$$

In particular, spin-$^3/_2$ fermions yield an axial anomaly 3 times that of spin-$^1/_2$ fermions in the case of Yang-Mills, and 3 - 24 = -21 times in the case of gravity. As an aside, we note that the relative size of the Yang-Mills and gravitational contributions, which at first sight seem unrelated, can be checked by topological arguments. The integrated version of Eq. (4.5) yields the index theorem

$$n_+ - n_- = \left[-\frac{s}{2} + s^3\right] \frac{1}{48\pi^2} \int d^4x \sqrt{g}\, R^*{}_{\mu\nu\rho\sigma} R^{\mu\nu\rho\sigma}$$

$$+ s \frac{e^2}{8\pi^2} \int d^4x \sqrt{g}\, F^*{}_{\mu\nu} F^{\mu\nu} \ , \quad (4.6)$$

where $n_+$ and $n_-$ are the number of right- and left-handed zero modes of the arbitrary spin Dirac operator (minus ghosts for $s > {}^1/_2$) and for simplicity we consider an Abelian gauge field. Each term on the right of Eq. (4.6) is proportional to a topological invariant and normally each term is separately an integer. On spaces which do not admit of an ordinary spin structure, however, only their sum will be an integer (Hawking & Pope 1978 b; Pope 1980). For example, on $CP^2$

$$n_+ - n_- = \left[-\frac{s}{2} + s^3\right] + s\, e^2 \ , \quad (4.7)$$

where e = odd half-integer, in which case one finds, consistently, that the right-hand side is indeed an integer for all off half-integer spin. Indeed, the argument could be turned around to deduce the gravitational anomaly from knowledge of the gauge-field anomaly, but only up to an integer (e.g. in the case of spin-$^3/_2$, one could not distinguish the factor -21 from +3, at least not without further detailed knowledge of $CP^2$). A discussion of arbitrary spin axial anomalies may also be found in the papers of Römer (1979, 1981).

Note that the trace anomaly and β-function calculations always involved even powers of spin, whereas the axial results involve odd powers of spin. [Strangely enough, we have been unable as yet to find any application of the spin sum rules (Eq. (3.2)) when k is odd.]

If one now applies these results to supersymmetric Yang-Mills theories one finds: a) The conformal anomalies and the axial anomalies form a multiplet along with the anomaly in $\gamma^{\mu}S_{\mu}$, where $S_{\mu}$ is the supercurrent, as one expects they should (Ferrara & Zumino 1975). For a review, see Grisaru (1978). b) The anomalies vanish for $N > 2$. This merely reflects the vanishing β-function for $N > 2$ discussed in Section 3.

A naïve application of these results to extended supergravity theories (with the usual field representations) leads to a more curious state of affairs. One finds: a) The anomalies do not appear to form a supermultiplet (Christensen & Duff 1978 a); b) The anomalies do not vanish for $N > 2$. Rather one finds that for $N > 2$, the A coefficient is an integer $A = 3 - N$ (Christensen et al. 1980). See Table 7.

Before enlarging on these remarks, we first discuss antisymmetric tensor fields.

### 4.2 Antisymmetric tensors: quantum inequivalence

The results just quoted for the A coefficient in supergravity are valid provided one makes the (ostensibly innocent) assumptions that: a) the auxiliary fields are irrelevant, b) the spin-0 particles are described by scalar fields $\phi(x)$. However, one can show that the A coefficient depends not only on the spin but also on the choice of field representation (Duff & van Nieuwenhuizen 1980). Thus the gauge theory of a rank-2 antisymmetric tensor $\phi_{\mu\nu}$ (with one degree of freedom) differs from that of a real scalar $\phi$; and that of an antisymmetric rank-3 tensor $\phi_{\mu\nu\rho}$ (with 0 degrees of freedom) differs from nothing. One finds

$$A[\phi_{\mu\nu}] = A[\phi] + 1$$

$$A[\phi_{\mu\nu\rho}] = -2 \; , \qquad (4.8)$$

hence the values quoted in Table 5. The reasons for this quantum inequivalence have been discussed at length elsewhere (Duff 1981 b,c). The important point is that such representations can, in fact, occur either as auxiliary fields in N = 1 supergravity or as physical fields in the versions of extended supergravity obtained by dimensional reduction. [Incidentally, the study of higher-rank antisymmetric tensor fields in particle physics has a distinguished history. I am very grateful to Professor Kemmer for drawing my attention to his 1938 paper (Kemmer 1938). Another reference is Ogivestsky & Polubarinov (1967).] For example, the N = 8 theory obtained by dimensional reduction from N = 1 in d = 11, taking the extra dimensions to be a 7-torus yields (Cremmer & Julia 1979) $63\phi + 7\phi_{\mu\nu} + \phi_{\mu\nu\rho}$. Only after making (topologically non-trivial) duality transformations does one obtain the theory with 70 $\phi$ fields. The importance of this, as noted by Siegel (private communication) is that the A coefficient of −5 in Table 7 becomes, on using Eq. (4.8)

$$A' = -5 + 7 - 2 = 0 \ . \tag{4.9}$$

Indeed, one can arrange that the vanishing of the A coefficient for all N > 2 (Duff 1981 b,c) with the choice of representations such that

$$3 - N[\psi_\mu] + N[\phi_{\mu\nu}] - 2N[\phi_{\mu\nu\rho}] = 0 \ , \tag{4.10}$$

where $N[\psi_\mu]$, $N[\phi_{\mu\nu}]$, and $N[\phi_{\mu\nu\rho}]$ are the number of gravitini, rank-two and rank-three antisymmetric tensors, respectively. Nicolai & Townsend (1981) have shows that $N \geq 3$ supergravity theories obeying Eq. (4.10) may be constructed from three basic N = 3 multiplets, one of which contains the antisymmetric tensor gauge field.

The idea that anomalies vanish in supergravity for N > 2 just as in Yang-Mills is certainly appealing, but at this stage several important questions remain unanswered:

1) Why does the antisymmetric tensor version of N = 8 with only SO(7) symmetry seem to have better ultraviolet behaviour than the scalar version with $E_7 \times SU(8)$ symmetry?

2) In order to fulfil Eq. (4.10), does one consider only physical fields or should one also consider auxiliary fields? Even if we confine our attention to the traditional minimal formulation of N = 1 auxiliary fields (Stelle & West 1978; Ferrara & van Nieuwenhuizen 1978 a) there is still some ambiguity in how one does the algebra. One or both of the scalars S and P could be replaced by $\partial^\mu S_\mu$ or $\partial^\mu P_\mu$ (Stelle & West 1978; Ogivetsky & Sokatchev 1980). Setting $\phi_{\mu\nu\rho} = \varepsilon_{\mu\nu\rho\sigma}S^{\sigma}$, etc., then yields the gauge theory of the rank-three field discussed above, and hence different anomalies. This problem becomes even more severe for N > 2 because no-one yet knows what the auxiliary fields are, or even if they exist!

3) Having allowed for antisymmetric tensors, do the anomalies now form a multiplet? We have seen that the inclusion of such representations can change the gravitational conformal anomaly. Interestingly enough, they can also contribute to the gravitational axial anomaly (Dowker 1978; Nielsen et al. 1978) if they are described by a first-order Lagrangian but not, apparently, if they are described by a second-order Lagrangian, since one cannot construct triangle graphs from the available vertices. [It may seem odd that one should ever describe bosons with a first-order Lagrangian, but such kinetic terms do in fact appear in the Lagrangian when one uses a manifestly supersymmetric gauge-fixing procedure (Grisaru & Siegel 1981 a).] One simple illustration of this puzzle is provided by the Wess-Zumino scalar multiplet, which describes a spinor, a scalar, and a pseudoscalar. When coupled to supergravity, the induced anomalies form a multiplet. If one now swaps the pseudoscalar for a gauge antisymmetric tensor one obtains a new multiplet (Siegel 1979). The conformal anomaly is now different but the axial anomaly seems not to have changed!

4) The quantum inequivalence discussed above was obtained using a background field gauge (Duff & van Nieuwenhuizen 1980). What is one to make of Siegel's claim (Siegel 1980) that there exists another gauge choice, e.g. a flat-space gauge, in which the inequivalences never occur? This would lead to the $A'$ coefficients for $\phi_{\mu\nu}$ and $\phi_{\mu\nu\rho}$ quoted in the second column of Table 5.

5) Is it significant that one can arrange for vanishing anomalies for N > 2, even with the <u>usual</u> representations, by taking the $A'$ coefficient of Eq. (4.4) and invoking the spin sum rules?

6) Why do these $A'$ coefficients differ from the A coefficients by <u>integers</u>? See Table 5.

7) Why does the barrier for vanishing anomalies occur for N > 2 both in Yang-Mills and supergravity? What does this imply for the N > 2 auxiliary field quest?

A partial set of answers to these questions can be obtained by re-examining the problem in terms of N = 1 superfields, and using the Grisaru-Siegel superfield Feynman rules (Grisaru & Siegel 1981 a), to which we now turn.

### 4.3 Chiral superfields: vanishing anomalies for $N > 2$

In this subsection, which was written with the help of Marc Grisaru, I shall summarize some results to be described in more detail in a future publication (Duff et al. 1981). The starting point is the form of the on-shell background field $N = 1$ supergravity action to second order in the quantum fields, after fixing the gauge and including the Fadeev-Popov ghosts (Grisaru & Siegel 1981 a):

$$S = \int d^4x\, d^4\theta\, E^{-1} \Bigg[ -\frac{1}{2} H \,\Box\, H - \frac{1}{2} H^{\alpha\dot{\beta}} \left( W_{\alpha}{}^{\gamma\delta} D_{\gamma} H_{\delta\dot{\beta}} + \bar{W}_{\dot{\beta}}{}^{\dot{\gamma}\dot{\delta}} \bar{D}_{\dot{\gamma}} H_{\alpha\dot{\delta}} \right) - \frac{18}{5} X\bar{X} + \sum_{i=1}^{3} \bar{\Psi}_{i}{}^{\dot{\beta}} i D_{\alpha\dot{\beta}} \Psi_{i}{}^{\alpha} + \sum_{i=1}^{2} \left( \Phi_{i}{}^{\alpha} D^2 \Phi_{i\alpha} + \text{h.c.} \right) \Bigg], \tag{4.11}$$

where the physical axial-vector superfields $H_{\alpha\dot{\alpha}}$ is real, the general spinors $\Psi_{\alpha i}$ have abnormal (ghost) statistics, the chiral spinors $\Phi_{\alpha i}$ have normal (ghost-for-ghost) statistics, and the chiral scalar X is a physical compensating field. The on-shell background supergravity fields appear in the determinant of the superveilbein E and in $W_{\alpha\beta\gamma}$ of Eq. (2.9). This action corresponds in components to the usual (S and P) auxiliary field choice. For N-extended supergravity one has to add contributions from superfields representing the $\{3/2, 1\}$, $\{1, 1/2\}$, and $\{1/2, 0\}$ matter multiplets, taking into account gauge-fixing and ghosts where necessary. The number of superfields of each type is given in Table 7 (V is a real scalar superfield) where the minus signs denote abnormal statistics. Again, the representation content is the usual one, e.g. $N = 8$ has 70 scalars.

To calculate the contribution to the one-loop anomalies from each superfield one may either compute a one-loop supergraph using the Feynman rules for superfields, or else infer it from the known component results. In the latter case, it is first necessary to replace the component contributions of Table 5, which correspond to physical fields after ghost subtraction, by those of Table 6 which are valid before ghost subtractions have been made. The remarkable result, as shown in Table 6, is that only chiral superfields contribute to anomalies. [This result may

Table 6

The A and A′ coefficients for both component fields and superfields after gauge fixing but before ghost subtraction. Note that (as in Yang-Mills) only chiral superfields yield anomalies.

COMPONENTS:

| | 360A | 360A′ | $\Delta A$ |
|---|---|---|---|
| $e_{\mu a}$ | 760 | -320 | 3 |
| $\psi_\mu$ | -212 | 148 | -1 |
| $\phi_\mu = \phi_{\mu\nu\rho}$ | - 44 | - 44 | 0 |
| $\chi$ | 7 | 7 | 0 |
| $\phi$ | 4 | 4 | 0 |
| $\phi_{\mu\nu}$ | 264 | - 96 | 1 |

N = 1 SUPERFIELDS:

| | 0 | A′ | $\Delta A$ | CHIRAL? |
|---|---|---|---|---|
| $H_a = e_{\mu a} + 4\psi_\mu + 4\phi_\mu + \phi_{\mu\nu}$ | 0 | 0 | 0 | NO |
| $\Psi_\alpha = \psi_\mu + 2\phi_\mu + 4\chi + 2\phi + \phi_{\mu\nu}$ | 0 | 0 | 0 | NO |
| $V = \phi_\mu + 4\chi + 4\phi$ | 0 | 0 | 0 | NO |
| $X = \chi + 2\phi$ | $\frac{1}{24}$ | $\frac{1}{24}$ | 0 | YES |
| $\Phi_\alpha = 4\chi + 2\phi + \phi_{\mu\nu}$ | $\frac{5}{6}$ | $-\frac{1}{6}$ | 1 | YES |

Table 7

Extended supergravities in terms of N = 1 superfields corresponding to the usual assignment of physical and auxiliary fields.

| N | H | ψ | V | X | ϕ | A | A′ | ΔA |
|---|---|---|---|---|---|---|---|---|
| 1 | 1 | -3 | 0 | 1 | 2 | $^{41}/_{24}$ | $-^{7}/_{24}$ | 2 |
| 2 | 1 | -2 | -1 | 2 | 1 | $-^{1}/_{12}$ | $^{11}/_{12}$ | 1 |
| 3 | 1 | -1 | -1 | 0 | 0 | 0 | 0 | 1 |
| 4 | 1 | 0 | 0 | -4 | -1 | -1 | 0 | -1 |
| 5 | 1 | 1 | 2 | -8 | -2 | -2 | 0 | -2 |
| 6 | 1 | 2 | 6 | -12 | -3 | -3 | 0 | -3 |
| 7 | 1 | 4 | 14 | -20 | -5 | -5 | 0 | -5 |
| 8 | 1 | 4 | 14 | -20 | -5 | -5 | 0 | -5 |

Table 8

A choice of N = 1 superfields yielding A = A′. The absence of chiral superfields for N > 2 yields vanishing anomalies.

| N | H | ψ | V | X | ϕ | A = A′ |
|---|---|---|---|---|---|---|
| 1 | 1 | -3 | 4 | -7 | 0 | $-^{7}/_{24}$ |
| 2 | 1 | -2 | 1 | -2 | 0 | $-^{1}/_{12}$ |
| 3 | 1 | -1 | -1 | 0 | 0 | 0 |
| 4 | 1 | 0 | -2 | 0 | 0 | 0 |
| 5 | 1 | 1 | -2 | 0 | 0 | 0 |
| 6 | 1 | 2 | 0 | 0 | 0 | 0 |
| 7 | 1 | 4 | 4 | 0 | 0 | 0 |
| 8 | 1 | 4 | 4 | 0 | 0 | 0 |

be verified directly from the superspace Feynman graphs by counting powers of $\theta$ in the vertices and propagators (see the lectures by Grisaru in this volume).] With these representations, therefore, one simply reproduces the results already quoted (Christensen & Duff 1978). In particular, A is an integer for $N > 2$.

However, an alternative action to Eq. (4.11), with a different choice of representations, can be obtained by employing a real scalar compensating field V, in which case (Grisaru & Siegel 1981 a):

$$S' = \int d^4x\, d^4\theta\, E^{-1} \left[ -\frac{1}{2} H \Box H - \frac{1}{2} H^{\alpha\dot{\beta}} \left( W_\alpha{}^{\gamma\delta} D_\gamma H_{\delta\dot{\beta}} + \bar{W}_{\dot{\beta}}{}^{\dot{\gamma}\dot{\delta}} \bar{D}_{\dot{\gamma}} H_{\alpha\dot{\delta}} \right) + \sum_{i=1}^{4} V_i \Box V_i + \sum_{i=1}^{3} \bar{\Psi}_i{}^{\dot{\beta}} i D_{\alpha\dot{\beta}} \Psi_i{}^\alpha + \sum_{i=1}^{7} X_i \bar{X}_i \right], \qquad (4.12)$$

where $\Psi$ and X have abnormal statistics. Note that the chiral spinor $\Phi_\alpha$ never appears. This action corresponds in components to exchanging one of the scalar auxiliary fields for a rank-three gauge antisymmetric tensor. With this starting point, one is led naturally to the superfield assignments of Table 8. One now finds that the components satisfy Eq. (4.10) and hence yield vanishing anomalies for $N > 2$. [Curiously, this seems to imply a connection between the $\phi_{\mu\nu\rho}$ field as an auxiliary field for $N = 1$ and the $\phi_{\mu\nu\rho}$ field obtained in $N = 8$ by dimensional reduction from $N = 1$ in $d = 11$.]

However, in terms of superfields, Table 8 yields a much simpler explanation for vanishing anomalies: The anomalies vanish for $N > 2$ because the net number of chiral superfields (i.e. physical minus ghost) is zero for $N > 2$. This provides another explanation, incidentally, for the vanishing one-loop $\beta$ function in $N > 2$ Yang-Mills. In terms of $N = 1$ superfields: $N = 1$ Yang-Mills requires one physical non-chiral field V and three ghost chiral fields X; $N = 2$ Yang-Mills requires one physical V, one physical X and three X ghosts; $N = 4$ Yang-Mills requires one physical V, three physical X, and three X ghosts. Hence the $\beta$ functions for $N = 1, 2, 4$ are in the ratio 3:2:0; the zero result for $N = 4$ being due to vanishing of the net number of chiral superfields.

In the absence of a complete formalism for N-extended superfields, we cannot give a fundamental explanation of why the net number of N = 1 chiral superfields should be zero for N > 2 in these models. The answer to this question may, however, lie in the results of Dr. Kallosh (this volume) in which she examines the consistency conditions for extended (as opposed to N = 1) superfields, and finds a discrete difference between $N \leq 2$ and N > 2 (due in the case of supergravity to the presence of spin-½ components for N = 3 onwards). Her results imply: i) no extended chiral matter superfields can exist in an extended superfield supergravity background for N > 2 (which might suggest that there should be no net number of N = 1 chiral superfields in any correct N = 1 analysis of N > 2 theories using the background field method and hence no anomalies for N > 2). This would certainly be consistent with her second claim: ii) no super extension of the Gauss-Bonnet invariant exists for N > 2.

### 4.4 Zero modes, gauge fixing, and boundary conditions

We have still not given an explanation for the different between the A and A′ coefficients. Let us first consider the fields $\phi_{\mu\nu}$ and $\phi_{\mu\nu\rho}$. By using a background covariant gauge Duff and van Nieuwenhuizen (1980) found "quantum inequivalence of different field representations", see Eq. (4.8). In a subsequent paper "Quantum equivalence of different field representations" Siegel (1980) found, using a different gauge

$$A'[\phi_{\mu\nu}] = A'[\phi]$$

$$A'[\phi_{\mu\nu\rho}] = 0 \; . \tag{4.13}$$

On the basis of this, Siegel concluded that the trace anomalies are gauge dependent and hence not physically observable. Here we give a different interpretation. A fuller explanation will be given elsewhere and here we summarize the salient points.

As explained in Duff and van Nieuwenhuizen (1980) and in Duff (1981 c), the reason for the quantum inequivalence was a topological one. Consider a p-form A, its exterior derivative dA and an action functional

$$S = \frac{1}{2}\,(dA, dA) \; . \tag{4.14}$$

S is invariant under the symmetry

$$A \to A + V \ , \tag{4.15}$$

where

$$dV = 0 \ . \tag{4.16}$$

In other words,

$$A \to A + V_H + d\alpha \ , \tag{4.17}$$

where $V_H$ is harmonic

$$\delta V = dV = 0 \ ; \qquad \delta = {}^*d^* \ . \tag{4.18}$$

Now consider the gauge-fixing addition $S'$

$$S' = \frac{1}{2} (\delta A, \delta A) \ . \tag{4.19}$$

The new action, given by

$$S + S' = \frac{1}{2} (A, \Delta A) \ ; \qquad \Delta = d\delta + \delta d \ , \tag{4.20}$$

although no longer invariant under the usual "small" gauge transformations $A \to A + d\alpha$, is still invariant under the "large" gauge transformations $A \to A + V_H$. The number of harmonic p-forms $V_H$ will be the number of zero eigenvalue modes of the corresponding Laplacian $\Delta_p$ which in a compact manifold is given by the Betti numbers, $n_p$. These are related to the Euler number Eq. (3.61) by

$$\chi = \sum_{p=0}^{4} (-1)^p \, n_p \ . \tag{4.21}$$

It is these zero modes in the functional integral which are responsible for the quantum inequivalence; the ghosts and ghosts-for-ghosts, etc., enter with alternating signs in the trace anomaly as in Eq. (4.21), yielding a coefficient of $\chi$ which differes by an integer from that expected on the grounds of naïve equivalence.

Suppose, on the other hand, one had chosen a gauge-fixing addition which broke both small and large gauge invariances. These zero-modes would now be absent and one would recover "quantum equivalence". This is our explanation for Siegel's result. Question: Which is correct? Answer: It depends on the boundary conditions!

The following simple example nicely illustrates the problem (and I am grateful to G.W. Gibbons, C.J. Isham, M. Peskin & E. Witten for discussions in it). Consider free Maxwell theory at finite temperature, i.e. on a manifold $M_3 \times S^1$ where $M_3$ is three-dimensional space. In a covariant gauge like Eq. (4.19) the partition function is given by

$$Z = \frac{\det \Delta_0}{(\det \Delta_1)^{1/2}} , \tag{4.22}$$

where $\Delta_1$ is the Laplacian on 1-forms (the physical potential $A_\mu$) and $\Delta_0$ is the Laplacian on 0-forms (the two scalar ghosts). However the operator $\Delta_1$ has a zero-mode ($S^1$ gives rise to a non-vanishing first Betti number, i.e. the space is not simply connected) corresponding to the field configuration

$$\begin{aligned} A_0 &= c = \text{constant} \\ &= \Omega\, i\partial_0 \Omega^{-1} , \end{aligned} \tag{4.23}$$

where

$$\Omega = \exp ict . \tag{4.24}$$

Now Eq. (4.23) is simply pure gauge and in Minkowski space $R^4$ would be trivial. At finite temperature, however,

$$\frac{1}{2\pi} \int_{S^1} A_\mu \, dx^\mu = n \tag{4.25}$$

[$n$ = integer if $\Omega \subset U(1)$] is non zero. Suppose, on the other hand, one had chosen the gauge

$$A_0 = 0 \tag{4.26}$$

one would automatically have ruled out all the non-trivial $n \neq 0$ configurations. We thus (apparently) would reach different partition functions in different gauges. Including charged matter via an $A_\mu J^\mu$ coupling would lead to a delta-function of total charge Q

$$Q = \int d^3x \, J^0 \tag{4.27}$$

in the gauge where one integrates over the zero mode but not in the gauge where the zero mode is omitted. Again one asks: which is correct? The answer depends on $M_3$. If, as is usual, $M_3$ is chosen to be just flat Euclidean space $R^3$ then Eq. (4.23) is not square integrable and this mode is omitted from the functional integral. If, however, $M_3$ were chosen to be compact, say $S^3$, then Eq. (4.23) would be square integrable and should be included. The lesson is that the answer is not gauge-dependent *per se*. Rather it is theory dependent: one obtains different answers for different physical boundary conditions. [It is simply that for a given boundary condition, there are good and bad gauges. For example Eq. (4.26) is a forbidden gauge in the compact case.] So it is with the antisymmetric tensors: one can have either "quantum inequivalence" or "quantum equivalence" depending on the physics. We emphasize, however, that for a given theory (i.e. Lagrangian plus boundary conditions) the trace anomaly is unique and gauge independent. For example, if the space-time is compact [as in Hawking's space-time foam approach (Hawking 1978)] one must obtain the A coefficient of Table 5, whereas the A′ coefficient would be appropriate for asymptotically flat space-times. It seems that similar remarks apply to spin-$^3/_2$ and spin-2: one can have pure gauge configurations

$$h_{\mu\nu} = \nabla_\mu \xi_\nu + \nabla_\nu \xi_\mu$$

$$\psi_\mu = D_\mu \varepsilon \tag{4.28}$$

which are still non-trivial. The omission of zero-modes when they cease to be square integrable also explains why $\Delta A$ is positive for commuting fields and negative for anticommuting. [Pure gauge modes of this kind have been discussed before in a somewhat different context by Hawking & Pope (1978 a). See also a recent paper by Gibbons (1981).] This newly found realization of boundary-condition dependence in anomaly calculations may also help to clear up some murky areas in the literature. For example, a Maxwell gauge field $A_\mu$ has zero degrees of freedom in two dimensions. By extrapolating the dimensional regularization (and flat-space gauge-fixing) calculations of Capper et al. (1974) to d = 2 one finds a vanishing trace for its one-loop quantum stress tensor (Duff 1977). The heat kernel method using a background field gauge, however, would yield

$$T^\mu{}_\mu = -\frac{1}{4\pi} R \tag{4.29}$$

or

$$\int d^4x \sqrt{g}\, T^\mu{}_\mu = -\chi , \tag{4.30}$$

where $\chi$ is the two dimensional Euler number. Indeed this example is precisely the d = 2 analogue of $\phi_{\mu\nu\rho}$ in d = 4. The whole debate about quantum inequivalence could have taken place years ago. In fact, it did. The coefficient that governs the R term in $T^\mu{}_\mu$ for d = 2 is the one that governs the $\Box R$ term in $T^\mu{}_\mu$ for d = 4. And there is a long-standing (and until now unresolved) debate about the coefficient of $\Box R$ in the trace anomaly for photons. See Duff (1977) for a list of references. [There is a somewhat different sense, of course, in which boundary conditions are important for anomalies. In space-times with boundary, the anomalies and counterterms will receive boundary term corrections. As explained in Duff (1981 c), moreover, these also cancel in supersymmetric theories obeying Eq. (4.10). See also Fradkin and Tseytlin (1981 a).]

In summary, therefore, one can obtain different anomalies in supergravity according to the choice of (a) physical field representations (b) auxiliary field representations, and (c) boundary conditions. For example, Table 7 shows that with suitable boundary conditions, the anomalies vanish for N > 2 even with the <u>usual</u> representations (thus restoring the $E_7 \times SU(8)$ symmetric version to its former glory). But what about that choice of fields and/or boundary conditions for which the anomalies do not vanish for N > 2 and for which, moreover, they do not appear to form a multiplet? It seems to me that, in this case, the zero-modes are actually breaking the supersymmetry (i.e. symmetry breaking via gravitational instantons). The argument which says anomalies <u>must</u> form a multiplet if one calculates with superfields now needs modification: the supersymmetric gauge-fixing inherent in the superfield formalism may simply be incompatible with the boundary or periodicity conditions. Similar phenomena are, in fact, already familiar from somewhat different contexts. For example, finite temperature for which the bosons are periodic in imaginary time but fermions antiperiodic is not compatible with rigid supersymmetry. Similarly, the vanishing of the vacuum energy density in supersymmetry (Zumino 1975) does not imply zero Casimir effect since perfectly conducting plates would place different boundary conditions on different components of a supermultiplet.

Finally, for those who have not found this section too confusing already, a few more words on auxiliary fields. Although we have discussed different representations in this section on anomalies, they each correspond to the old minimal set (Stelle & West 1978; Ferrara & van Nieuwenhuizen 1978 a). In other words, they correspond to $n = -1/3$ in the language of Gates et al. (1981). However, these authors have recently examined the "new-minimal" auxiliary fields of Sohnius and West (1981 b) corresponding to $n = 0$, as well as the old non-minimal set of Breitenlohner corresponding to $n \neq 0, -1/3$, in the context of anomalies [in this case the coupling to matter is via the "new deteriorated energy momentum tensor" (Duff & Townsend 1981)] and find the presence of trace anomalies will in general lead to anomalies in the Ward identities for local supersymmetry. They conclude that for N = 1 supergravity, at least, only the $n = -1/3$ set lead to a consistent theory.

We close this section on anomalies with another unanswered question. At one loop we have learned that no chiral superfields means no trace anomalies and hence, by supersymmetry, no chiral anomalies either. This is a very satisfactory result. Chiral anomalies rely on an imbalance between left- and right-handed fields and so such imbalance is possible in the absence of chiral superfields. Is this merely a one-loop phenomenon, or does it have ramifications for the exact theory?

## 5 RECENT DEVELOPMENTS

Since the bulk of these lectures were written, there have been several interesting developments in our understanding of the ultraviolet divergences in extended supergravity. The most important are summarized below.

### Sum rules

Although the spin sum rules amd mass sum rules discussed in Section 3 may be verified by inspection of the mass and spin assignments of extended supersymmetry multiplets, it would obviously be advantageous to derive these relations from first principles. Such a derivation has now been given by Ferrara et al. (1981), starting from the fundamental algebra of extended supersymmetry. Indeed this work confirms one's suspicions that the spin and mass rules have the same origin. Moreover, they find that the rules used to demonstrate one-loop ultraviolet divergence cancellations are merely special cases of more general rules which are suggestive of new, as yet undiscovered, cancellation phenomena. See also Ferrara (1981).

### N = 8 supergravity with local SO(8) × SU(8)

As discussed in Section 3, the vanishing of the one-loop $\beta$-function in gauged N > 4 supergravity was discovered before the Lagrangians for these models had been completely constructed and hence before their actual existence was established. Now, De Wit & Nicolai (1981 a,b) have succeeded not only in constructing gauged N = 8 (and hence by truncation N < 8) but in exhibiting a local, non-linearly realized, SU(8) invariance in addition to the local linear SO(8). Their assignments are

| | $e_\mu{}^a$ | $\psi_\mu{}^i$ | $A_\mu{}^{IJ}$ | $\chi^{ijk}$ | $U_{ij}{}^{IJ}$ | $V^{ijIJ}$ |
|---|---|---|---|---|---|---|
| SO(8) | 1 | 1 | 28 | 1 | 28 | 28 |
| SU(8) | 1 | 8 | 1 | 56 | $\overline{28}$ | 28 |

[One may recover the anticipated 1, 8, 28, 56, and 70 representations of SO(8) by imposing a special SU(8) gauge.] As one switches off the gauge coupling one recovers the version of ungauged N = 8 with $E_7$(rigid) × × SU(8) (local) invariance (Cremmer & Julia 1979).

This gauged model of N = 8 offers different possibilities for superunification from the ungauged theory. One possibility, suggested by De Wit & Nicolai (1981 a,b) is that only the SO(8) singlets of spin 2, $^3/_2$, and $^1/_2$ are unconfined, though it remains unclear whether this is compatible with the vanishing β-function. An analysis of possible counterterms in this model has been given by Howe & Nicolai (1981).

## Conformal supergravity: finite for N = 4

Although fourth-order theories were mentioned in the Introduction, we have not yet discussed those encountered in supergravity, for example, the conformal supergravities with N = 1, 2, 3, and 4, which are the supersymmetric extensions of the (Weyl tensor)$^2$ conformal gravity. [For a review, see De Wit (1981).] One would not expect such theories to be consistent at the quantum level because the existence of conformal anomalies at one loop would force non-conformal modifications to the Lagrangian at two loops and beyond (Capper & Duff 1975), and the conformal invariance is necessary to maintain the number of propagating degrees of freedom. This objection could be avoided, of course, if the theory were ultraviolet finite. Recently, Fradkin & Tseytlin (1981 a,b,c) have calculated the one-loop counterterms for N = 1, 2, 3, and 4 and found that N = 4 is one-loop finite! This is yet another example of a "miraculous" cancellation not expected on the grounds of invariants-as-counterterm arguments.

## Topologically massive gravity

Whilst on the subject of higher derivatives, we mention the recent work of Deser et al. (1981) who consider a three-dimensional gravity theory with a third derivative "topological" mass term. This term is enough to yield power counting renormalizability; yet surprisingly, in contrast to most higher derivative theories, yields no ghosts. A supersymmetric version no doubt exists also, but the implications for four-dimensional supergravity, if any, remain unknown.

## Non-renormalization theorems: proofs of finiteness to all orders

We have emphasized in Section 3 the inadequacy of the absence-of-invariants-as-counterterms arguments in explaining divergence cancellations in theories with extended supersymmetry. Useful though the knowledge of higher derivative invariants may be, this line of thought may even be seen ultimately to have delayed progress in supergravity by several years!

It was already known in the early days of global supersymmetry that the ultraviolet behaviour of supersymmetric theories was better than might be expected from naïve power counting, even for N = 1. For example, the absence of quadratic mass renormalization for scalar partners of chiral fermions in the Wess-Zumino model was known some time ago (Wess & Zumino 1974 a; Illiopoulos & Zumino 1974). Similar non-renormalization theorems occur in Abelian (Wess & Zumino 1974 b) and non-Abelian (Ferrara & Piguet 1975) gauge theories. This latter phenomenon has stimulated much of the recent interest in supersymmetric GUTS. See, for example, Ferrara (1981).

These non-renormalization phenomena are but special cases of an even stronger result proved by Grisaru et al. (1979) using N = 1 superspace Feynman rules: all quantum corrections to the effective action can be written as a single $\int d^4\theta$ integral over the full N = 1 superspace. This means that terms in the tree Lagrangian which cannot be so written (e.g. the mass and self interaction terms for chiral scalar superfields, $\int d^2\theta\ \phi^n$, in the Wess-Zumino model) will never be renormalized!

However, the far-reaching implications of this result have only recently become fully appreciated. In a recent paper Stelle (1981) exploits the presence of such an $\int d^2\theta$ chiral interaction in the N = 4 Yang-Mills theory written in terms of N = 1 superfields and then invokes the fourfold supersymmetry to prove that, since this term is non-renormalized, all terms are non-renormalized. Thus the β-function remains zero to all orders of perturbation theory! It also appears that this same N = 1 superfield non-renormalization theorem can be invoked to prove the vanishing of the β-function to all orders in gauged supergravity for N > 4 (Stelle & Townsend 1981).

## String theories

The historical foundations of supersymmetry are intimately tied up with the dual string. Recently there have been two interesting, though separate, developments in this connection. The first, due to Polyakov (1981 a,b) relates the critical dimensions of d = 26 for the bosonic string and d = 10 for the fermionic string to anomalies in the trace of the stress tensor (Section 4) in two-dimensional gravity and supergravity theories, respectively. It seems that the topological and boundary corrections to the anomalies discussed in Section 4 will also be important, when one considers possible holes and boundaries for the two-dimensionsal sheet swept out by the string. After these trace anomalies are taken into account, the resulting theory is described by an interacting Liouville Lagrangian, the mass parameter being related to the cosmological constant which Polyakov introduces on the grounds of renormalizability. In the author's opinion, incidentally, no such interaction is required for the fermionic string because no cosmological term is induced by quantum corrections. Indeed, this is a trivial example of the non-renormalization theorem discussed above. The absence of an induced cosmological term has been confirmed in explicit calculations by Fradkin & Tseytlin (1981 d).

A second interesting development, due to Green et al. (1981) concerns the emergence of N = 1 Yang-Mills in d = 10 and N = 2 supergravity in d = 10 as the zero-slope limits of the fermionic string models. By considering the extra dimensions to be compactified as circles and then letting the radii of the circles tend to zero, they are able to analyse the ultraviolet and infrared divergecnes of the theories for different values of d. Interestingly, they find that at one loop both theories are ultraviolet (UV) finite for $d < 8$ and infrared (IR) finite for $d > 4$, the IR divergence being milder in the gravitational case. They speculate that if certain kinematical features persist at higher loops the N = 4 Yang-Mills in d = 4 will be UV finite to all orders but N = 8 supergravity would diverge at three loops and beyond. Interestingly, their results for two-, three-, and four-particle Green's function and thus speculations on the UV behaviour of these theories for various values of N exactly parallel those in Section 3 made on the basis of the sum rules and the rough guide of at most one power of spin at each vertex for Yang-Mills at most two powers of spin for gravity.

These authors also stress, of course, that the original string theories might yet be UV finite even if the zero-slope limits are not.

## More supergraphs

Grisaru & Siegel (1981 b) have continued to exploit their Feynman rules for N = 1 superfields. A recent application is the calculation of the one-loop four-particle S matrix in extended supergravity. They find the calculations hardest for N = 1 supergravity, becoming progressively easier with higher N culminating in an almost trivial result for N = 8, in agreement with the above-mentioned string calculations. The absence of chiral superfields for $N \geq 3$ discussed in Section 4 considerably simplifies the results, and the vanishing of various terms in the amplitudes for increasing N again strongly suggests that the spin sum rules of Section 3 are at work.

[We have not discussed in these lectures the theory of "gauge supersymmetry" due to Arnowitt and Nath, since the particle content is difficult to extract and in any event includes unphysical ghost states. However, these authors were among the first to exploit the power of superspace power counting. Indeed, their theory, albeit unphysical, can be shown to be finite to all orders. See, for example Arnowitt & Nath (1979). Further discussions of UV divergences and superfields may be found in Taylor (1979).]

In a more recent paper on superspace power counting rules Grisaru & Siegel (1982) are able (subject to the existence of extended superfield formalism): a) to confirm the finiteness of N = 4 Yang-Mills to all orders, b) to extend the vanishing two-loop contribution to $\beta$ for N = 2 (Section 3) to all loops greater than one, c) to rule out counterterms in N = 8 supergravity in the first six loop orders (this means, in particular, that the suggested three one-loop counterterm is not in fact present).

At this stage, one might regard the present inability to rule out all counterterms for N = 8 as a disappointment, but time and again we have learned never to underestimate the power of supersymmetry!

Kaluza-Klein

One possibility remaining is the one suggested in the Introduction: perhaps N = 1 supergravity is finite in d = 11 (Duff & Toms 1981). Incidentally, the Brink-Green-Schwarz results do not include this possibility. They consider a theory which corresponds to N = 2 in d = 10 and this is obtained from N = 1 in d = 11 only after discarding an infinite tower of massive states. [An analysis of the kinds of massive states which arise in Kaluza-Klein theories has recently been given by Salam & Strathdee (1981). Interestingly, they find that they belong to infinite dimensional representations of non-compact groups.]

Although Cremmer & Julia (1979) were aware of the possibility of choosing the extra seven dimensions to be something other than the 7-torus ($S^1 \times S^1 \times \ldots S^1$), they restricted their explicit calculations to this case. In this way, they obtained an N = 8 supergravity in d = 4 with a rigid SO(7) invariance. (This is the antisymmetric tensor version discussed in Section 4.) Only after performing duality transformations (which convert antisymmetric tensors to scalars) did they obtain the version with $E_7$(rigid) × SU(8)(local). The geometric origin of these symmetries is therefore obscured.

However, in a forthcoming paper (Duff & Pope 1982) it will be shown that a gauged version of N = 8 may be obtained from N = 1 in d = 11, but by taking the extra dimensions to be a 7-sphere ($S^7$), which guarantees that the four-dimensional Lagrangian for the massless states will have a local SO(8) invariance and a (negative) cosmological constant. The resulting theory is therefore almost certainly that of De Wit & Nicolai (1981) discussed above. Further confirmation comes from a counting of the massless states: we find 1 graviton, 8 gravitini, 28 vectors and 56 spin-½ fermions. A direct count of the number of massless scalars is rather more difficult and is presently under way, but indirectly one can argue that the number of physical scalars must be 70 owing to the 8-fold supersymmetry, which in turn is guaranteed by the presence of eight massless spin-$^3/_2$ particles. What can be said with certainty is that they are genuine scalars and not antisymmetric tensors. (This is due to the difference in Betti numbers between a 7-sphere and a 7-torus.) Consequently no duality transformations are required to obtain the symmetries of the De Wit-Nicolai Lagrangian. It seems likely therefore that the hidden SU(8) (local) in the gauged version and the hidden $E_7$(rigid) one obtains on switching off the gauge coupling (shrinking the sphere to zero radius) may have a more transparent geometrical explanation in the picture with the 7-sphere than in the picture with the 7-torus.

This Kaluza-Klein framework puts into perspective the difference between the two N = 8 theories discussed in Section 4. Indeed, there could be an infinite number of d = 4 theories corresponding to the infinite number of ways of compactifying the extra seven dimensions, though N = 8 supersymmetry (eight massless gravitini) would be the exception rather than the rule. The question of ground-state stability still remains open but this large number of different theories might even be reinterpreted as many different phases of the same theory. Some of them would display the SU(3) × SU(2) × U(1) symmetry of the Witten (1981) scheme. The crucial question is still the number of fermions. Rules for counting the numbers of massless states in Kaluza-Klein supergravity will be given elsewhere (Duff & Pope 1982).

Can the vanishing β-functions of Section 3 and the vanishing anomalies of Section 4 be understood from higher dimensions? What is the connection between the non-renormalization theorems discussed above, the spin and mass sum rules of Section 3, and the absence of chiral superfields of Section 4? Can we find off-shell formulations of extended supergravity in d = 4, or N = 1 in d = 11 and, if so, how will they affect the renormalization?

## Conclusion

Our understanding of the ultraviolet divergences in extended supergravity is only just beginning.